\documentclass[twoside,twocolumn,9pt]{article}
\usepackage{extsizes}
\usepackage[super,sort&compress,comma]{natbib} 
\usepackage[version=3]{mhchem}
\usepackage[left=1.5cm, right=1.5cm, top=1.785cm, bottom=2.0cm]{geometry}
\usepackage{balance}
\usepackage{mathptmx}
\usepackage{sectsty}
\usepackage{graphicx} 
\usepackage{lastpage}
\usepackage[format=plain,justification=justified,singlelinecheck=false,font={stretch=1.125,small,sf},labelfont=bf,labelsep=space]{caption}
\usepackage{float}
\usepackage{fancyhdr}
\usepackage{fnpos}
\usepackage[english]{babel}
\addto{\captionsenglish}{%
  
}
\usepackage{array}
\usepackage{droidsans}
\usepackage{charter}
\usepackage[T1]{fontenc}
\usepackage[usenames,dvipsnames]{xcolor}
\usepackage{setspace}
\usepackage[compact]{titlesec}
\usepackage{hyperref}

\usepackage{epstopdf}

\definecolor{cream}{RGB}{222,217,201}

\begin{document}

\pagestyle{fancy}
\thispagestyle{plain}
\fancypagestyle{plain}{
\renewcommand{\headrulewidth}{0pt}
}

\makeFNbottom
\makeatletter
\renewcommand\LARGE{\@setfontsize\LARGE{15pt}{17}}
\renewcommand\Large{\@setfontsize\Large{12pt}{14}}
\renewcommand\large{\@setfontsize\large{10pt}{12}}
\renewcommand\footnotesize{\@setfontsize\footnotesize{7pt}{10}}
\makeatother

\renewcommand{\thefootnote}{\fnsymbol{footnote}}
\renewcommand\footnoterule{\vspace*{1pt}%
\color{cream}\hrule width 3.5in height 0.4pt \color{black}\vspace*{5pt}} 
\setcounter{secnumdepth}{5}

\makeatletter 
\renewcommand\@biblabel[1]{#1}            
\renewcommand\@makefntext[1]%
{\noindent\makebox[0pt][r]{\@thefnmark\,}#1}
\makeatother 
\renewcommand{\figurename}{\small{Fig.}~}
\sectionfont{\sffamily\Large}
\subsectionfont{\normalsize}
\subsubsectionfont{\bf}
\setstretch{1.125} 
\setlength{\skip\footins}{0.8cm}
\setlength{\footnotesep}{0.25cm}
\setlength{\jot}{10pt}
\titlespacing*{\section}{0pt}{4pt}{4pt}
\titlespacing*{\subsection}{0pt}{15pt}{1pt}

\fancyfoot{}
\fancyfoot[RO]{\footnotesize{\sffamily{1--\pageref{LastPage} ~\textbar  \hspace{2pt}\thepage}}}
\fancyfoot[LE]{\footnotesize{\sffamily{\thepage~\textbar\hspace{3.45cm} 1--\pageref{LastPage}}}}
\fancyhead{}
\renewcommand{\headrulewidth}{0pt} 
\renewcommand{\footrulewidth}{0pt}
\setlength{\arrayrulewidth}{1pt}
\setlength{\columnsep}{6.5mm}
\setlength\bibsep{1pt}

\makeatletter 
\newlength{\figrulesep} 
\setlength{\figrulesep}{0.5\textfloatsep} 

\newcommand{\topfigrule}{\vspace*{-1pt}%
\noindent{\color{cream}\rule[-\figrulesep]{\columnwidth}{1.5pt}} }

\newcommand{\botfigrule}{\vspace*{-2pt}%
\noindent{\color{cream}\rule[\figrulesep]{\columnwidth}{1.5pt}} }

\newcommand{\dblfigrule}{\vspace*{-1pt}%
\noindent{\color{cream}\rule[-\figrulesep]{\textwidth}{1.5pt}} }

\makeatother

\twocolumn[
  \begin{@twocolumnfalse}
\vspace{1em}
\sffamily
\begin{tabular}{m{0cm} p{17.5cm} }

& \noindent \LARGE{\textbf{Drop Impact on Hot Plates: Contact times, Lift-off and the Lamella Rupture}} \\
\vspace{0.3cm} & \vspace{0.3cm} \\

& \noindent \large{Sang-Hyeon Lee,$^{\ddag}$\textit{$^{a}$} Kirsten Harth,$^{\ddag \ast}$\textit{$^{b,c}$} Maaike Rump,\textit{$^{c}$} Minwoo Kim,\textit{$^a$} Detlef Lohse,\textit{$^{c,d}$} Kamel Fezzaa,\textit{$^e$} and Jong Ho Je\textit{$^{\ast a}$}} \\
\vspace{0.3cm} & \vspace{0.3cm} \\

& \noindent \normalsize{When a liquid drop impacts on a heated substrate, it can remain deposited, or violently boil in contact, or lift off with or without ever touching the surface. The latter is known as the Leidenfrost effect. The duration and area of the liquid--substrate contact is highly relevant for the heat transfer, as well as other effects such as corrosion.
However, most experimental studies rely on side view imaging to determine contact times, and those are often mixed with
the time until the drop lifts off from the substrate. Here, we develop and validate a reliable method of contact time
determination using high-speed X-ray and Total Internal Reflection measurements. We exemplarily compare contact and
lift-off times on flat silicon and sapphire substrates. We show that drops can rebound even without formation of a
complete vapor layer, with a wide range of lift-off times. On sapphire, we find a local minimum of lift-off times much
shorter than by capillary rebound in the comparatively low-temperature regime of transition boiling / thermal atomization.
We elucidate the underlying mechanism related to spontaneous rupture of the lamella and receding of the contact area.} \\

\end{tabular}

 \end{@twocolumnfalse} \vspace{0.6cm}
  ]

\renewcommand*\rmdefault{bch}\normalfont\upshape
\rmfamily
\section*{}
\vspace{-1cm}


\footnotetext{\textit{$^{a}$~ X-ray Imaging Center, Department of Materials Science and Engineering, Pohang University of Science and Technology, 77 Chengam-Ro, Nam-Gu, Pohang, 37673, Repbulic of Korea. E-mail: jhje@postech.ac.kr}}
\footnotetext{\textit{$^{b}$~ Institute of Physics, Otto von Guericke University Magdeburg, Universit\"atsplatz 2, 39106 Magdeburg, Germany. E-mail: kirsten.harth@ovgu.de}}
\footnotetext{\textit{$^{c}$~Physics of Fluids, Max Planck Center and University of Twente, Drienerlolaan 5, 7500NB Enschede, The
        Netherlands.}}
\footnotetext{\textit{$^{d}$~ Max Planck Institute for Dynamics and Self-Organization, Am Fa\ss{}berg 17, 37077 G\"ottingen, Germany. }}
\footnotetext{\textit{$^{e}$~ X-ray Science Division, Advanced Photon Source, Argonne National Laboratory, 9700 South Cass Avenue Argonne, Illinois 60439, USA.}}
\footnotetext{\ddag~ These authors contributed equally to this work and share first authorship of this article.}


\section{Introduction}
The impact of liquid drops on a heated substrate can be easily
observed in a hot pan during cooking: If the pan is hot enough,
droplets will levitate on a layer of their own vapor, move seemingly
frictionless and evaporate only very slowly. Upon impact from some
initial height, they will even rebound and jump several
times. 
At lower surface temperatures, the drops touch the substrate, and
display violent boiling behavior with numerous bubbles and spray.
Such droplets will evaporate in much shorter time than the
levitating ones. The original description of these phenomena goes
back to Boerhave~\cite{Boerhave1732} and
Leidenfrost~\cite{Leidenfrost1756,Wares1966}, see also the review by
D. Quere~\cite{Quere2013}. They have since been observed in many
everyday and engineering situations, e.g., in spray cooling of hot
metals or electronics~\cite{Hall1995,Kim2007,Liang2017}, in spray
combustion~\cite{Panao2005} or in the cooling of fuel rods in
nuclear power plants in case of an accident~\cite{Hamdan2015}. In
such situations, controlling their deposition and the heat transfer
between the solid and the droplets is crucial, because improper
droplet behavior can lead to material failure or to increased
corrosion. Besides material parameters, the heat transfer between
the substrate and the drop is determined mainly by the duration and
extent of the contact between the liquid and the solid surface.


Drops which are gently deposited on a hot smooth surface show an
immediate transition from contact boiling to film boiling at the
static Leidenfrost point $T_{L,s}$. This transition temperature is
influenced by parameters of the liquid and of the solid, e.g.,
accounting for the rate of evaporation and for cooling effects. The
loss of contact between the substrate and the liquid at $T_{L,s}$ is
accompanied by an immediate increase in the droplet's lifetime, and
vice versa a decrease in the heat transfer rate away from the
substrate to the droplet~\cite{Biance2003}. The shape of such \emph{static} Leidenfrost drops is
essentially determined by a balance of capillary and gravitational
forces. Typically, an upward vapor bulge forms near the symmetry
axis at the drop's underside~\cite{Snoeijer2009,Sobac2014,Burton2012}.
When the drop exceeds a critical size, this shape becomes unstable leading to
lamella rupture and the upward ejection of vapor bubbles~\cite{Biance2003}. Even solid objects
can display the Leidenfrost effect when approaching a hot plate
slowly enough~\cite{Burton2013,Waitukaitis2018}.

The situation gets more complicated when a drop impacts onto the hot
surface with some significant velocity. To distinguish it from the
usually lower \emph{static} Leidenfrost temperature of gently
deposited drops, this was termed the \emph{dynamic} Leidenfrost
effect~\cite{Tran2012}. Here, the \emph{transition boiling} regime
arises between full contact and complete levitation. In this case,
the radially outward region of the lamella loses contact with the
substrate and levitates while the central region retains
contact~\cite{Shirota2016}. When the substrate cooling through the
interaction with the drop is strong, all regimes may be observed
during a single impact event~\cite{Limbeek2016}.

Another complication arising from the impact situation is the
dynamics of the vapor layer. First, a thin layer of the ambient gas
is entrained between any impacting drop and a smooth surface at
ambient pressure~\cite{Kolinski2012,Kolinski2014}. This allows a
contact-less rebound even on smooth non-heated substrates for very
low Weber numbers, ${\rm We} = \rho
U_0^2D_D/\gamma<5$~\cite{Kolinski2014} with $U_0$ the impact
velocity, $D_D$ the drop diameter, $\rho$ the density and $\gamma$
the surface tension of the liquid. The drop's bottom deforms to a
dimple upon approach, leading to initial contact along a
ring~\cite{Mandre2009,Duchemin2011,Bouwhuis2012}. On a hot surface,
vapor is produced additionally, so that the velocity dependence of the transition temperature to film boiling, the {\em dynamic}
Leidenfrost point $T_{L,d}$, is a priori not clear~\cite{Tran2012}.
It cannot be determined from side view imaging, frequently applied
in the literature. High-Speed Frustrated Total Internal Reflection
(TIR) imaging allowed the determination of the {\em dynamic}
Leidenfrost transition, and yields increasing $T_{L,d}$ with
increasing initial substrate temperature. Measurements were based on
the initial approach of the drop~\cite{Shirota2016,Limbeek2016}. The
same technique can also elucidate the time-dependent contact
morphology, e.g., revealing different kinds of fingering
structures~\cite{Khavari2015,Harth2019} and oscillatory wetting
states~\cite{Khavari2017,Harth2019}, or solidification in vicinity of a cold plate~\cite{Kant2020,Koldeweij2019}. Complementary to that, one can
determine the height profile of the liquid/vapor contact line
projected onto a 2D plane from Ultrafast X-ray phase contrast
imaging ~\cite{Lee2018a,Lee2012,Lee2018,Lee2020}. From this, the
existence of contact is not obvious due to limited spatial
resolution. Our new method employing X-ray refraction overcomes this
problem.

The time scale that a droplet resides in the vicinity of a hot plate
(the \emph{lift-off} time) depends on numerous factors and it ranges
from a few milliseconds to seconds depending on the boiling
scenarios~\cite{Bernardin1997,Breitenbach2017}. The literature is not
concise about those scenarios, usually classified on the basis
of spray formation, rebound and decomposition detected in
conventional side and top view imaging is employed, see e.g.
Refs~\cite{Tran2012,Bertola2015,Staat2015,Liang2016,Clavijo2017}.
Drop rebound was observed even in contact
situations~\cite{Bertola2015,Roisman2018}. The formation of spray
drops and other fragments induced by the rupture of vapor bubbles
through the lamella is summarized under the term  ''secondary
atomization'' ~\cite{Cossali2005,Roisman2018}. Spray formation due
to bubbles bursting through the lamella is expected at the time when
the thickness of the thermal boundary layer reaches the local
lamella thickness~\cite{Roisman2018}. Expanding macroscopic holes
can induce breakup into comparatively large
fragments~\cite{Biance2011,Breitenbach2017,Roisman2018}.

Here, hole formation can be thermally induced by lamella rupture in the
contact region and independent of $\rm{We}$~\cite{Roisman2018}. The lamella rupture in particular does not require the introduction of topographical
defects on the substrate as in
Refs.~\cite{Biance2011,Chantelot2018}. For general aspects
related to rupture and fragmentation of thin fluid films see the
review by Villermaux~\cite{Villermaux2007} and, e.g.,
Refs.\cite{Taylor1959,Culick1960,Kooij2018,Neel2018,Poulain2018,Wang2018}. The rebound of spread-out
droplets before retraction (pancake rebound) reduces the contact
time with superhydrophobic substrates~\cite{Chantelot2018,Liu2014}.
Due to their large solid-liquid contact angles and low contact-angle
hysteresis, drop spreading and receding behavior is in some respect
expected to be similar to Leidenfrost drops.

 Contact and lift-off
times of drops impacting on hot substrates have been only
insufficiently discriminated in the literature due to the lack of
proper measurement techniques. For water, contact times measured
from side and bottom views~\cite{Breitenbach2017,Roisman2018} were
reported to be roughly independent of the substrate temperature and
$\rm We$ in the range of $T_s=\left[210...290\right]^{\circ}$C
(regime of secondary atomization), but scaling as ${\rm
Re}_0^{-4/5}$ with the Reynolds number ${\rm Re_0}=\rho_0 U_0
D_D/\eta_0$ at ambient temperature, where $\eta_0$ is the dynamic
viscosity. Liang et al.~\cite{Liang2016} report residence times
(from side views) of ethanol and butanol drops independent of the
surface temperature in the range of $T_s=\left[200 \dots 400
\right]^{\circ}$C for $\rm{We}<100$ (without details of the boiling
regime), which are in good agreement with the correlations from
Biance et al.~\cite{Biance2006}, Chen et al.~\cite{Chen2007}, and
their own suggested empirical relation, which all roughly follow a
scaling $\propto {\rm We}^{0.5}$. {\color{black} The residence times,
boiling and atomization regimes are also affected by the wettability
of the substrate~\cite{Clavijo2017}, with atomization preferentially
present on low-contact-angle substrates. However, in particular
substrates with very high contact angles are usually not flat, with
the topography introducing additional complexity for, e.g., heat
transfer, vapour flow and  bubble dynamics.}

Our interest in this study is a {\color{black} more} reliable
determination of the contact time and the lift-off or rebound
characteristics of impacting drops on well heat-conducting hot
smooth substrates. TIR can only be employed on transparent
substrates, i.e., metal surfaces are excluded. X-ray projection
imaging using the phase and absorption contrast of synchrotron
X-rays in interaction with the drop can visualize the vapor
structures within the liquid droplets independent of the substrate's
transparency. Our new approach using X-ray refraction data,
validated against TIR measurements below, provides access to contact
times of impacting drops. Moreover, we relate details of the contact
morphology to conventional side and top view videos. We distinguish
four impact regimes by the lift-off, contact and spray formation
characteristics, and relate these transitions to the droplet's
dynamics and to the lift-off times.

We first explain the experimental setup
(Sec.~\ref{Sec:Experimental}). Then, in Sec.~\ref{Sec:RegContTimes},
we provide a classification of contact / impact regimes, introduce
and validate our new method to measure the contact times from X-ray
refraction and last determine contact and lift-off times as well as
processes on smooth silicon and sapphire. Sec.~\ref{Sec:Holes}
addresses the surprising appearance of a local minimum of contact
times in the transition boiling regime, whose physical background we
subsequently explain. This paper ends with conclusions and an
outlook (Sec.~\ref{Sec:Conclusions}).

\section{Experimental Methods}
\label{Sec:Experimental}

\begin{figure}[ht]
        \centering
        \includegraphics[width=\columnwidth]{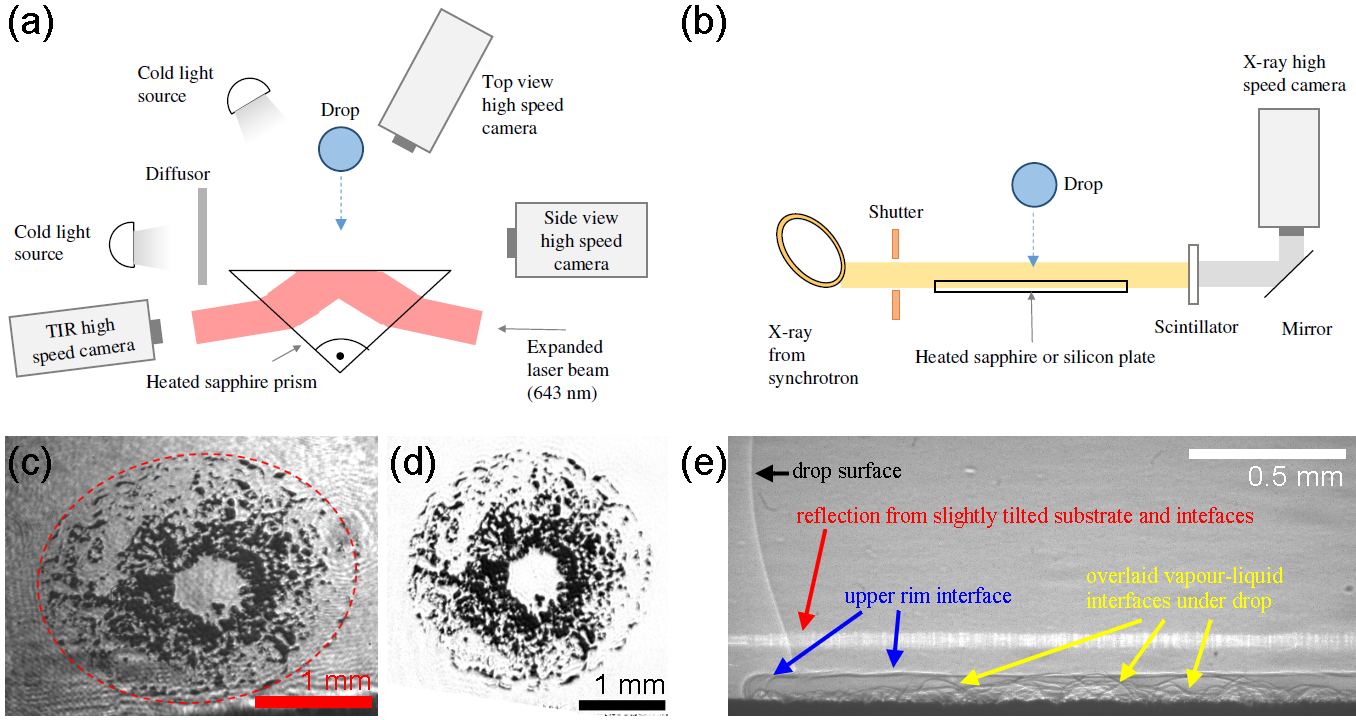}
        \caption{Schematic of the experimental setups for (a) high-speed TIR combined with synchronized side and top view imaging and (b) ultrafast phase contrast X-ray imaging. Example images show (c) an original frame of the TIR data, (d) the corresponding processed TIR image (see text) and (e) an X-ray image with annotations regarding the visible interfaces and reflections. The contact morphology and local thickness of the vapour layer is highly dynamic and inhomogeneous, as seen by the distribution of the contacting (black) spots in (c,d) and the profile in (e). The data show a $R=1~\rm mm$ ethanol drop impacting at $U\approx 1~\rm m/s$ on a sapphire substrate heated to $T_s=160^{\circ}$C at $t=0.48~\rm ms$ (c,d) and $t=0.31~\rm ms$ (e) after impact. The dashed line in (c) indicates the rotated, elliptical appearance of the drop's footprint due to the optical method.
\label{fig:setup}}
    \end{figure}

We study ethanol drops impacting on bulk optically smooth silicon
disks (from Thor Labs in X-ray, thickness 2 mm) and sapphire
substrates (disks of 2 or 3 mm thickness obtained from Edmund Optics
and Crystan Ltd. in the X-ray experiments, sapphire prism from
Crystan Ltd. for TIR). The setups together with example images are
shown in Fig.~\ref{fig:setup}. The motivation for using these
substrates is avoiding substrate cooling during the interaction with
the drop as much as possible, given for the silicon window, and the
need of transparency for the
direct observation of the contact dynamics using TIR.\\

\subsection{General Aspects}

We study the impact of ethanol droplets of diameter $2~{\rm mm}\leq
D_D\leq2.1~\rm mm$ from heights between 4 and 10 cm, corresponding
to low impact velocities between $\approx 0.88~\rm m/s$ and $1.4~\rm
m/s$. Droplets were produced by pinch-off due to a slow volume
increase using a syringe pump. If not noted differently, we used
steel needles of standard gauge 21G and 22G (inner diameter 0.5 and
0.6 mm, outer diameters 0.8 and 0.9 mm, respectively). The
temperature of the droplet was not controlled. However, in
particular at increased temperatures we discarded a few drops prior
to the actual experiment, in order to avoid assembled liquid
contamination due to evaporation. The Reynolds and Weber numbers for
a $D_D=2~\rm mm$ diameter drop impacting at $U_0=1~\rm m/s$ (the
standard case for most experiments reported here) are ${\rm
Re}=\rho_0 U_0 D_D/\eta_0 \approx 1380$ and ${\rm We}=\rho_0 U_0^2
D_D/\gamma_0 = 71.2$, respectively, using the values of the dynamic
viscosity $\eta_0 = 1.14\cdot 10^{-3}~\rm Pa\,
s$~\cite{EthDynViscosity}, density $\rho_0=786.9 ~\rm kg
m^{-3}$~\cite{EthDensity} and surfac tension $\gamma_0 = 22.1 \cdot
10^{-3}~\rm N m^{-1}$~\cite{Muratov1980} of ethanol at room
temperature ($22^{\circ}$C). These parameters change to $\eta_b =
0.42\cdot 10^{-3}~\rm Pa\, s$~\cite{EthDynViscosity},
$\rho_{b}=727.9 ~\rm kg m^{-3}$~\cite{EthDensity} and $\gamma_b =
17.1 \cdot 10^{-3}~\rm N m^{-1}$~\cite{Muratov1980} at the boiling
point of ethanol, $T_b=78.3 ^{\circ}$C, thus ${\rm Re}_b=3466$ and
${\rm We}_b=85.2$. The latent heat of vaporization and the isobaric
specific heat capacity are $L=841~\rm kJ\,
kg^{-1}$~\cite{Matyushov1994} and $c_p=3.182~\rm kJ\, kg^{-1}
K^{-1}$~\cite{EthHeatCap}, respectively. We use the values at room temperature in
the following, as the actual temperature due to heating of the drop has not been measured while it was hanging from the needle .

{\color{black} Both substrates are partially wetting for ethanol at
room temperature. For ethanol on untreated silicon, one can
extrapolate to a static contact angle of $14.5\pm 3^{\circ}$ from
Ref.~\cite{Spencer2013}. We expect a similar contact behavior for
ethanol on sapphire. After receding, a sessile drop forms a
spherical cap of
contact angle $22\pm 2^{\circ}$. Note that dynamic and static contact angles in non-isothermal situations are usually strongly affected by thermal effects.}\\

The droplets directly impact onto the silicon or sapphire
substrates. Given temperatures refer to the calibrated initial
surface temperature $T_s$. The thickness of the thermal boundary
layer in the substrate can be estimated from the scaling
$d_{\rm th}\propto \sqrt{\alpha t}$~\cite{Limbeek2016}, where $t$ is the
duration of substrate--liquid contact. The silicon window has a
thermal
diffusivity of $\alpha_{Si}\approx 0.4\cdot 10^{-4}\rm m^2
s^{-1}$~\cite{Shanks1963} at $180^{\circ}$C. For the sapphire
substrates, $k_{s,Sa} = 36 ~\rm W m^{-1} K^{-1}$ at
$180^{\circ}$C~\cite{Touloukian1967}, $\rho_{Sa}=3960~\rm
kg\,m^{-3}$ at 180$^{\circ}$C \footnote{The density was estimated
using the density value at room temperature, $3970~\rm kg\,
m^{-3}$~\cite{SapphProp}, and assuming weighted linear thermal
expansion so that $V(T)=V(20^{\circ}{\rm
C})\left[1+\beta_{\parallel}(T-20^{\circ}{\rm
C})\right]\left[1+\beta_{\perp}(T-20^{\circ}{\rm C})\right]^2$ with
the linear coefficients of expansion $\beta_{\parallel} = 5.6\cdot
10^{-6} ~\rm ^{\circ}$C and $\beta_{\perp} = 5.0\cdot 10^{-6} ~\rm
^{\circ}$C ~\cite{SapphProp} }, $C_{p,Sa} = 101.7~\rm J\, mol^{-1}
K^{-1}$ at $180^{\circ}$C~\cite{Ditmars1981}, $m_{mol,Sa}=101.96~\rm
g\, mol^{-1}$, leading to a thermal diffusivity $\alpha_{Sa}=
k_{s,Sa} m_{mol,Sa}/(\rho_{Sa}C_{p,Sa})\approx 10^{-5}~\rm m^2
s^{-1}$. The resulting thickness of the thermal boundary layer,
after our time scale of drop spreading of 3.5 ms at the given $\rm
Re$ and $\rm We$, is $\approx 22~\rm \mu m$ in silicon and $\approx
10~\rm \mu m$ in sapphire. Thus, we assume that substrate cooling is
negligible in our experiment, similar to Ref.~\cite{Limbeek2016}.

In the X-ray experiments, $T_s$ was varied between
$T_s=100^{\circ}$C and $600^{\circ}$C with $1\%$ accuracy to cover
all boiling regimes from contact to film boiling using an electronic
heater (SU-200-IH, Maivac). It was directly connected to a thermocouple
(UNI-T UT 325) placed in the copper block, and $T_s$ was calibrated separately on the
substrate's surface using a PT100 probe. The substrates were
directly placed on the copper block of the heater. The setup is as
described in Refs.~\cite{Lee2018,Lee2020}.

The TIR experiments were performed separately. In this case the
substrate was a right-angle sapphire prism displaying a top surface
of $25 \times 35.35 ~\rm mm^2$ placed in a custom-made heated
aluminum block. The surface temperature could be controlled between
room temperature and $358^{\circ}$C with an accuracy of $\approx
1.5~\rm K$ using a proportional-integral-derivative controlled
electrical heating system. The surface temperature was directly
measured and calibrated against the heater set-point using a PT100
temperature sensor flat attached to the surface. The general setup
and TIR method have been previously described in
Ref.~\cite{Shirota2017}.

\subsection{Imaging Techniques}
\textbf{Ultrafast synchrotron X-ray imaging} were conducted at
the 32-ID undulator beamline of the Advanced Photon Source, Argonne
National Laboratory, to achieve high temporal and spatial
resolutions, with a 472 ns exposure time for each frame. Using the
setup shown in Fig.~\ref{fig:setup}(b), we are capable of directly
visualizing the interfaces between gas and liquid even within the
liquid drop at a maximum time resolution of up to $\approx 271,000$
frames per second (fps). The imaging system consists of a fast
scintillator and a mirror coupled to a high-speed camera (Photron
Fastcam SA-Z) via a microscope objective lens (10x with NA=0.21),
resulting in a spatial resolution of $2~\rm \mu m / pixel$. Typical
frame rates used in the present measurements are between $20,889$
and $90,000$ fps. As the setup had to be remotely controlled, a
laser triggering system was used to sense the falling drop and to
trigger the camera and the X-ray beamline shutter. The method has
been successfully applied previously to drops impacting on
non-heated solid substrates and onto liquid
layers~\cite{Zhang2011,Zhang2012,Lee2015} as well as heated
substrates~\cite{Lee2018,Lee2018a,Lee2020}. Heating of the
scintillator due to the intense radiation limits the total
measurement duration to $\approx 40~\rm ms$.

The strength of edge-enhanced x-ray phase contrast imaging is the
possibility to visualize both the drop liquid and the vapor phase as
demonstrated in Fig.~\ref{fig:setup}(e), and the ability to detect
in particular even small distortions of liquid-gas interfaces
irrespective of substrate transparency. Measurements were performed
in transmission, so that all interfaces in the direction of the beam
are overlaid on a single image. An example of an ethanol drop
impacting onto a sapphire disk heated to $T_s=160^{\circ}$C (surface
slightly tilted toward the incoming X-ray beam, {\color{black} tilt
angle $<0.1^{\circ}$}) is shown in Fig.~\ref{fig:setup}(e), and the
appearance of the gas-liquid interfaces and the substrate is
clarified. For the later analysis, the {\color{black} substrate was
positioned horizontally, by adjusting the tilt so that the
reflection barely disappears at the substrate. Minimal tilt
corresponds to the minimal height at which the black substrate
region appears in absence of the drop. The refraction in the
substrate region remains and will be analyzed.} Due to the limit of
the spatial resolution and the typically immense number of bubbles
forming directly at the substrate in the transition boiling regime,
the existence of contact is not obvious. We developed a novel
approach using the structure of the X-ray refraction at the
gas-liquid interfaces underneath the drop (see below). Note that the
hard X-ray irradiation causes negligible heating and vaporization of
the liquid or change of the properties of liquids for the very short
exposures
used here ($<300~\rm \mu s$)~\cite{Weon2008}.\\

High-speed Frustrated \textbf{Total Internal Reflection (TIR)}
imaging relies on the fact that light is totally internally
reflected at the interface between an optically dense and an
optically dilute medium when exceeding a critical angle of
incidence. This allows a reliable detection of wetted regions on the
substrate, apart from temperatures close to the dynamic Leidenfrost
point when the contact features approach the limit of experimental
spatial and temporal resolution. 
With some limitations, also accurate quantitative height data in the range of the evanescent wave can be retrieved~\cite{Shirota2017}. Our TIR setup was combined
with synchronized side view imaging, and for part of the experiments
also tilted top view imaging, as sketched in
Fig.~\ref{fig:setup}(a).

For the TIR sketched in Fig.~\ref{fig:setup}(a), we introduce a s-polarized expanded continuous laser beam, $\lambda = 643~\rm nm$, of roughly 2 cm spot diameter into our sapphire prism (refractive index $n =1.76$~\cite{SapphProp}) at an angle such that it is totally internally reflected at the gas-sapphire interface, while it is transmitted into the ethanol drop in contact. The actual angle is not important for the current measurement, but can be calculated from the distortion of the recorded TIR images~\cite{Shirota2017}. Under these conditions, all spots wetted by the drop appear dark in the images. As seen in Fig.~\ref{fig:setup}(c), the original recorded images distort the circular drop footprint to an ellipse, whose axis is additionally rotated to the frame borders as we used a second mirror to guide the reflected beam to the camera lens. All images shown below are normalized by their background and undistorted, exemplarily shown in Fig.~\ref{fig:setup}(d). TIR recordings presented here were recorded at frame rates between 30,000 and 80,000 fps at a spatial resolution in the raw images of typically $7.45~\rm \mu m/pixel$, and few low temperature data sets at $12.3~\rm \mu m/pixel$.\\

In \textbf{side view}, we employed a cold light source and a
diffusor to provide backlight. For the \textbf{tilted top view}, a
cold light source was placed at a mirrored opposing position to the
camera lens, such that the light reflected on the prism's surface
provides a bright background. These recordings were employed for
extraction of the drop size, impact velocity, the lift-off or
rebound times (side) and to observe the hole formation morphology
(tilt view). The spatial and temporal resolution varies between data
sets, depending on the process of interest, we employed frame rates
between 4,000 and 30,000 fps, most side view recordings possess
spatial resolutions between $12~\rm \mu m/pixel$ and $20~\rm \mu
m/pixel$.

\section{Contact Regimes, Contact and Lift-off Time}
\label{Sec:RegContTimes}

First, we categorize our impacting drops into four main regimes
observed at an impact velocity of $U\approx 1~\rm m/s$, based on the
contact and the lift-off / rebound characteristics seen in the vapor
layer dynamics. In particular, the distribution of contact locations
on the substrate is by far not spatially homogeneous, see
Fig.~\ref{fig:Scenarios}. We define the dynamic Leidenfrost point as
the lowest temperature where no contact between the drop and the
substrate is observed during the entire spreading and receding
phase, which is identified by combination of TIR and X-ray, details
will be discussed elsewhere~\cite{Harth2020}. We find $T_{L,d}\approx
300^{\circ}$C. The actual dynamic Leidenfrost point may be slightly
{\color{black} ($10 ~ 20^{\circ}$C) lower under cleanroom
conditions.} Nevertheless, this is substantially higher than the
value of $200^{\circ}$C to $220^{\circ}$C based on the initial
contact alone reported by Shirota et al.~\cite{Shirota2016}, and
similarly by Khavari and Tran~\cite{Khavari2017}. In repeated
experiments, contact was occasionally also observed at temperatures
up to $340^{\circ}$C, with decreasing probability. A transitional
boiling regime is found in a broad range of temperatures between the
contact and Leidenfrost boiling
regimes~\cite{Shirota2016,Limbeek2016}. Then, the maximal radius of
the wetted area in TIR is equal to or smaller than the spreading
radius observed in top or side view, respectively. We observe a
lower limiting temperature for transition boiling of $(161\pm
2)^{\circ}$C. Oscillatory wetting~\cite{Khavari2017,Harth2019},
where radially inward traveling wetting fronts characterize the TIR
recordings, is already observed at slightly lower surface
temperatures of $156^{\circ}$C. As we are interested in the contact
and lift-off times, we will adopt a different classification here.

Second, we determine the contact and lift-off or rebound times of
ethanol drops on sapphire and silicon substrates. Atomization gives
rise to a local minimum of the lift-off time on sapphire. We
{\color{black} qualitatively} explain its relation to the receding
contact in the transition boiling regime and the localized rupture
of the lamella in Sec.~\ref{Sec:Holes}.

\subsection{Impact Scenarios}

\begin{figure}[ht]
        \centering
        \includegraphics[width=\columnwidth]{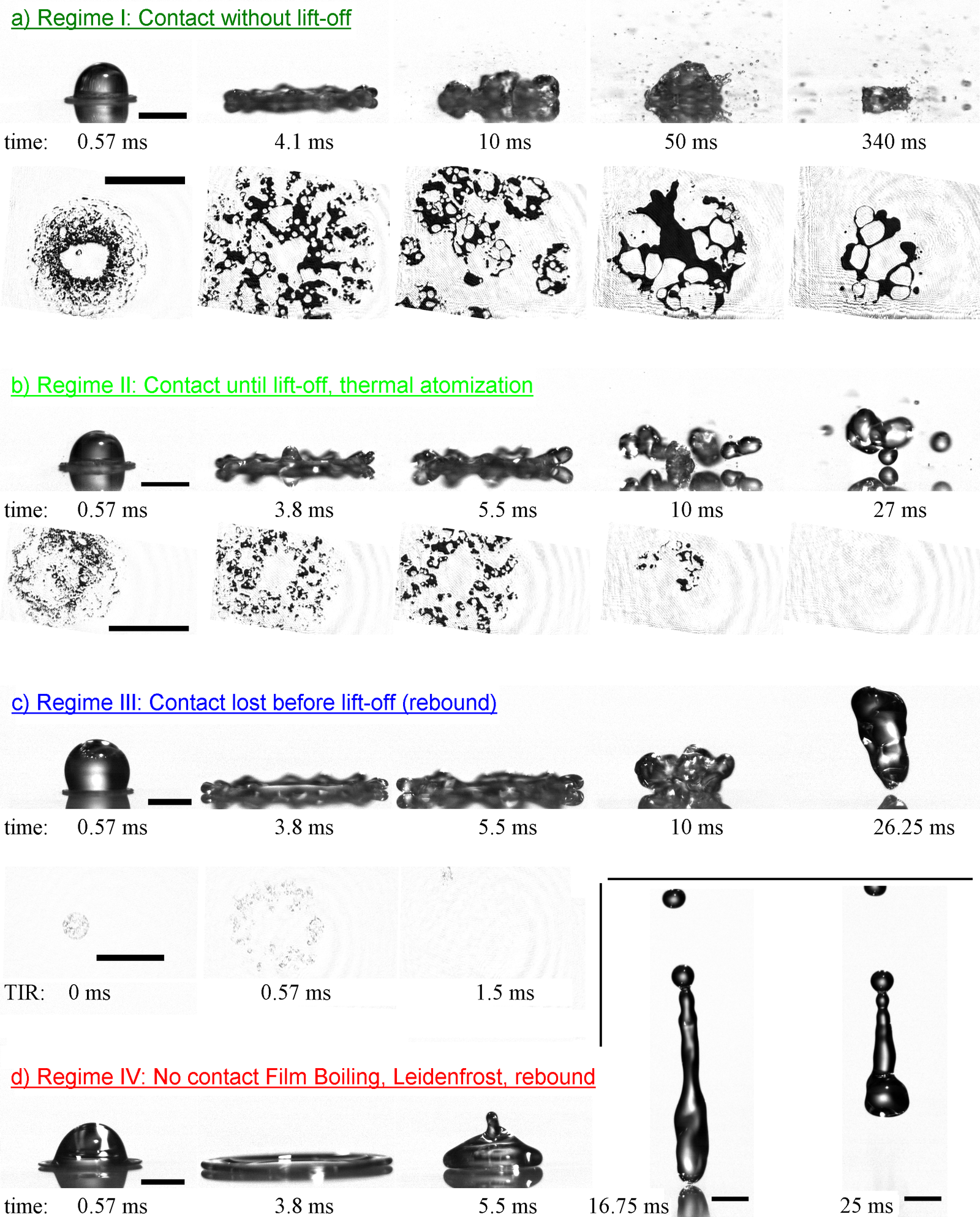}
        \caption{Contact and lift-off regimes in optical measurements, see corresponding Fig.~\ref{fig:ScenariosX} for
        X-ray: Ethanol, drop diameter 2 mm, impact velocity 1 m/s,
        $We=71.6$,
        initial sapphire substrate temperatures $T_s$ (a) $160^{\circ}$C, (b) $168^{\circ}$C, (c) $204^{\circ}$C and (d) $323^{\circ}$C.
        (a) Regime I - drops stick to the plate, accompanied with strong spray formation until complete evaporation.
        (b) Regime II - drops stick to the plate until the retraction releases the contact so that lift-off occurs (compare TIR and side view), usually accompanied by (upward and sideways) spray and sometimes drop fragmentation;
        (c) Regime III - contact vanishes (no TIR signal after 1.6 ms) long before the drop rebounds, (d) Regime IV - Leidenfrost drops, which never touch the substrate.
        Time is given after the first frame seen in TIR (a-c), and from contact in side view in (d) as there is no contact seen in
        TIR. See text for detailed regime description.
        Scale bars are 2 mm in each sub-figure. Note the slightly different scale between the time ranges in (d) $[0.57\dots 5.5]~\rm ms$ and (d) $[10,26.25]~\rm ms$.
        \label{fig:Scenarios}} 
\end{figure}

\begin{figure}[ht]
        \centering
        \includegraphics[width=\columnwidth]{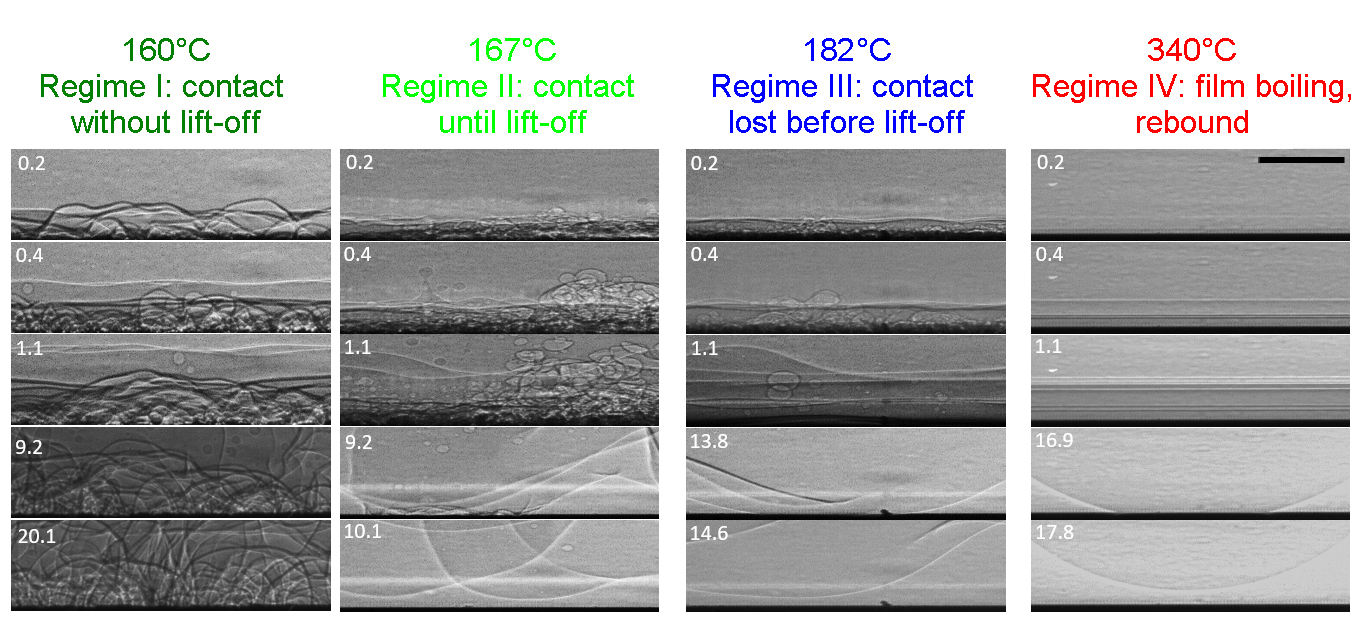}
        \caption{Contact and lift-off regimes of drops impacting on a smooth hot plate detected by side view X-ray
        imaging, description as in Fig.~\ref{fig:Scenarios}. Numbers give the time after impact in $~\rm ms$.
        Note the persisting contact until $t=9.2~\rm ms$ in Regime II, compared to the early loss of contact in Regime III.
        The length of the scale bar is 0.2 mm and applies to all images. The Weber number is 71.2, drop diameter 2 mm.  \label{fig:ScenariosX}}
\end{figure}

\begin{figure}[ht]
        \centering
        \includegraphics[width=\columnwidth]{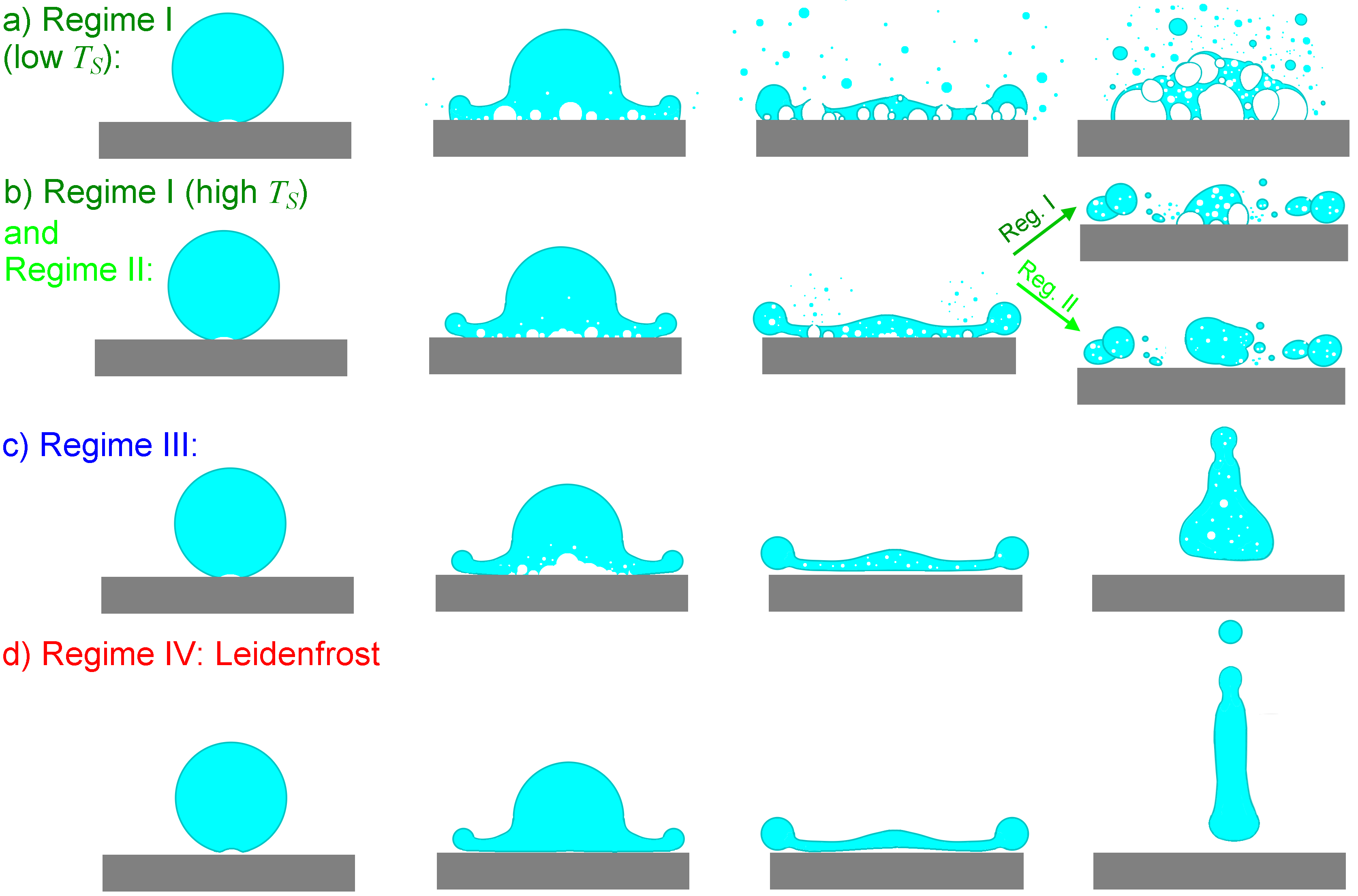}
        \caption{Sketch of the drop behaviour in the 4 impact regimes, Figs.~\ref{fig:Scenarios} and~\ref{fig:ScenariosX}.
        All cases are similar at impact (first column), vapour formation proceeds faster with increasing $T_S$, (a--d), and column
        4 shows the impact outcome after drop receding. \label{fig:ScenariosSketch}}
\end{figure}

The four distinguished impact and contact regimes with their
appearance in optical and X-ray data are exemplarily shown in
Figs.~\ref{fig:Scenarios} and~\ref{fig:ScenariosX} and schematically drawn in Fig.~\ref{fig:ScenariosSketch}. The scenarios are similar on sapphire and silicon substrates,
as well as for the entire investigated range of impact velocities.\\

Regime I, Figs.~\ref{fig:Scenarios}(a),~\ref{fig:ScenariosX}(a) and~\ref{fig:ScenariosSketch}(a),
corresponds to the contact boiling regime, where drops impact onto a substrate of $T_s$
sufficiently high above the boiling point, but at comparatively low
temperatures. They spread on the substrate and
start violently boiling. Numerous vapor bubbles nucleate, grow
and sometimes merge, and subsequently burst through the drop's
surface or sideways near the contact line of the drop on the
substrate, until the drop is completely decomposed into smaller
spray drops and the main drop evaporated on the substrate. 
The contact times are of the order of $0.5~\rm s$ at
$T_s=160^{\circ}$C in our experiment. The numerous moving contact
lines and often millimeter-sized
bubbles are easily identified in the optical and X-ray data.\\

In Regimes II-IV, the droplets lose contact with the substrate prior
to complete evaporation (identified by TIR and X-ray data). This can
occur by a classical rebound, where the drops jump to a significant
height. In particular at lower temperatures, the drops (or their
fragments) rise from the substrate at small velocity to a clearly
visible but small height (much less than $R_0$), see Fig.~\ref{fig:ScenariosSketch}(b). As we would like to
distinguish this from the classical rebound, we rather generalize
the term to \emph{'lift-off'}, including any of the two situations.

The lift-off and contact times, $t_l$ and $t_c$, are often not
cleanly distinguished in the literature, as this is impossible from
simple side and top view data. The literature often presents $t_c$
as the visible loss of contact with the substrate in side view or
the lack of visible structures on the lamella from top
view~\cite{Roisman2018}. One the one hand, this causes
an overestimation of $t_c$ in
Regimes II and III. On the other hand, the dynamic Leidenfrost
transition is underestimated due to very short and sparse contacts
at high temperatures~\cite{Harth2020}.

We extract the lift-off time, $t_l$, from the side views in phase
contrast X-ray and TIR data. We also determine the \emph{actual}
contact time, $t_c$, which is often much smaller than $t_l$ (see
below). Note that structures on the spread out lamella in optical
top view images may persist for a while after contact was already
lost, and are thus not an indication of persisting contact.

In Regime II, Figs.~\ref{fig:Scenarios}(b),~\ref{fig:ScenariosX}(b)
and~\ref{fig:ScenariosSketch}(b),part of the drop continuously
remains in contact with the substrate until lift-off. 
At a temperature of $T_s>165^{^\circ}$C (for ethanol on sapphire,
$D_D=2 ~\rm mm$, $U_0=1~\rm m/s$), we observe cases of drops
detaching from the substrate after substantially shorter contact
times than in Regime I, on the order of 10-20 ms. The number of
vapour structures substantially increases, while their average size
and lifetime decreases (seen from TIR, Fig.~\ref{fig:Scenarios}(b)
and particularly well in X-ray, Fig.~\ref{fig:ScenariosX}(b),
$t=0.2-1.1~\rm ms$). Often, the drop fragments in a spread-out
state, particularly well seen in top view (see
Fig.~\ref{fig:minimum_imageSeq}). Similar lift-off of drops after
contact with a hot substrate was observed
before~\cite{Breitenbach2017,Roisman2018}, but never analyzed even
qualitatively. Generally, drop bouncing from hot substrates was
regarded to be due to the formation of a complete vapor layer
between the drop and the substrate~\cite{Tran2012}, i.e. the dynamic
Leidenfrost transition. However, our experiments evidence than the
bare transformation of surface energy to kinetic energy upon drop
retraction together with a sufficiently rapid receding contact
region are enough to cause a drop to lift off the hot substrate,
even if the Leidenfrost transition is by far not reached.
{\color{black} As this small contact region is usually close to the
impact point,} we can clearly identify local, moving contact points
separated by vapor bubbles in the X-ray data of
Fig.~\ref{fig:ScenariosX}(b), $t=9.2~\rm ms$.  We carried out at
least 10 experiments per temperature, and we can observe both
sticking and lift-off in individual impact events between
$T_s=165^{\circ}$C and $168^{\circ}$C, cf. also
Sec.~\ref{Sec:Holes}. At lower temperatures, only sticking is
observed. This temperature range roughly coincides with the
emergence of the transition boiling regime. Significant spray
formation is characteristic for Regime II. The high-T range of
Regime I and Regime II (Fig.~\ref{fig:ScenariosSketch}(b))
correspond to the 'thermal
atomization' observed and described in Ref.~\cite{Roisman2018}.\\

Upon further increase of the substrate temperature, we enter Regime
III: Contact is lost before the drop lifts off the substrate, as
seen in Figs.~\ref{fig:Scenarios}(c),~\ref{fig:ScenariosX}(c) and
~\ref{fig:ScenariosSketch}(c). The central region of contact becomes
smaller and more porous. If spray is observed, it occurs sideways
during the early spreading phase.  Drops retract as a whole. In the
example, $t_l\approx26~\rm ms$, while the TIR data reveal a more
than ten times shorter contact time of only $t_c=1.6~\rm ms$. The
corresponding X-ray images, Fig.~\ref{fig:ScenariosX}(c), are
characterized by smooth, slowly evolving liquid-vapor interfaces
after contact is lost (shown in the image for $t=13.8~\rm ms$ at
$T_s=182^{\circ}$C). At high temperatures, approaching the dynamic
Leidenfrost point, we observe a slight delay of a few microseconds
between impact and initial contact in both TIR and X-ray data. We
suppose that this can be related both to substrate cooling similar
to the drops impacting on glass in Ref.~\cite{Limbeek2016} and to
the presence of small surface or liquid contamination due to dust.
In particular, there is a broad range of $\approx 40^{\circ}$C above
the dynamic Leidenfrost point, where short, delayed contact can be
observed in some realizations. Thus, for individual impact events,
we may observe regime III and the Leidenfrost regime IV at
the same substrate temperature.\\

In regime IV, the Leidenfrost or film boiling regime, the drop
continuously levitates on a vapor layer and never contacts the
substrate. In TIR, a spreading, grey-scale ring may be seen in
particular at lower temperatures, where the vapor layer is
sufficiently thin such that the evanescent wave is partially
transmitted (see Ref.~\cite{Shirota2017} for details). In the X-ray
data, the drop's bottom is perfectly smooth. As the drop spreads, a
vapor rim forms behind the droplet's rim, and capillary waves evolve
on the lamella during retraction~\cite{Lee2018,Lee2020}.
The drops rebound by contracting to a slender liquid column from
which several smaller satellite drops are usually ejected upwards.
In the example of Fig.~\ref{fig:Scenarios}(d), 6 satellites are
formed starting after $t=13~\rm ms$. Rebound occurs after
$t_l=16.75~\rm ms$. The maximum levitation of the drop's bottom is
$\approx~\rm 4.4~\rm mm$, reached after $44~\rm ms$ (Determination
of the height of the center of mass is complicated as the drop
decomposes and
parts of it leave the field of view.).\\

\begin{figure}[ht!]
        \centering
        \includegraphics[width=\columnwidth]{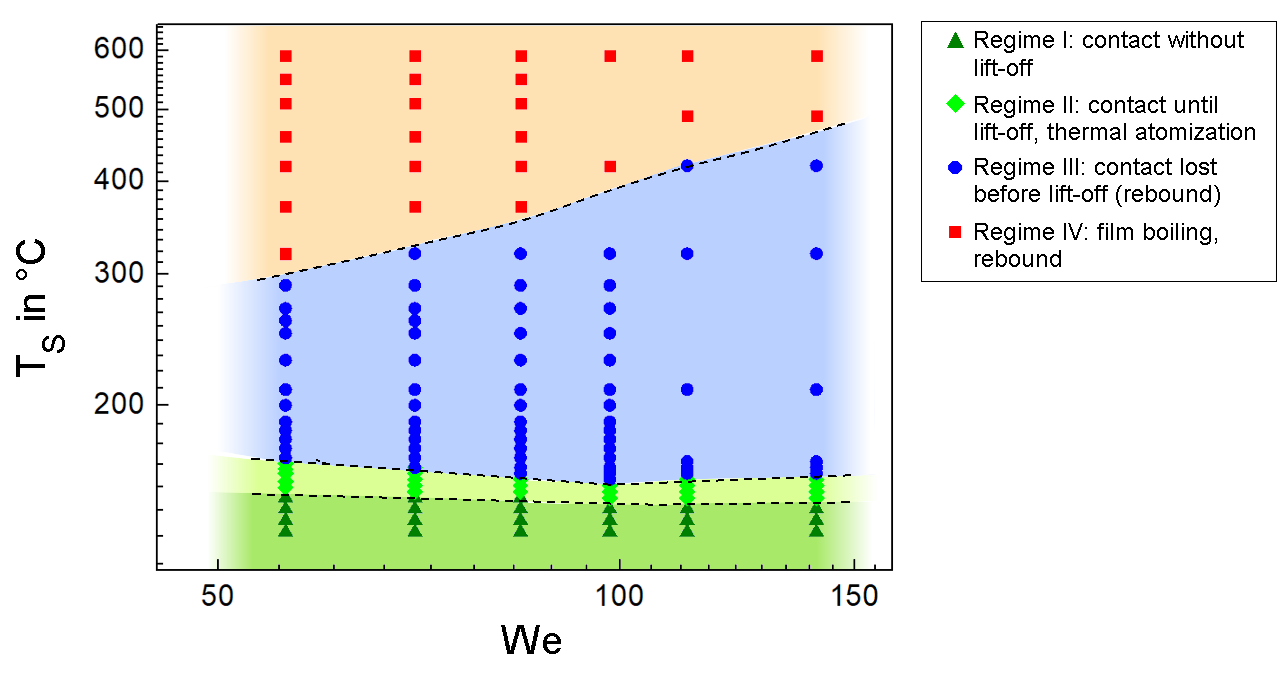}
        \caption{State diagram for ethanol drop impact on a heated silicon substrate with different $\rm We$ showing
        the four different impact regimes (see text and Figs.~\ref{fig:Scenarios},~\ref{fig:ScenariosX},~\ref{fig:ScenariosSketch}). Dashed lines are guides to the eye indicating the Regime transitions.}
\label{fig:StateDiagram}
\end{figure}

Finally, we report the occurrence of the respective regimes at
varied impact Weber number $\rm We$ extracted from X-ray experiments
of 2 mm diameter ethanol drops impacting on a silicon substrate in
the state diagram in Fig.~\ref{fig:StateDiagram}. The points are
deduced from 5 to 10 repeated experiments for each point. Only the
dynamic Leidenfrost temperature (transition  of regime III
$\rightarrow$ IV) strongly increases with increasing impact velocity
(Weber number), while the cross-over between the contact / lift-off
regimes display no (I $\rightarrow$ II, at $150^{\circ}$C) or little
(regime II $\rightarrow$ III, between 160$^{\circ}$C and
170$^{\circ}$C) temperature dependence within experimental accuracy.
The irrelevance of the Weber number (at constant drop size) for the
existence of regime II (similar to thermal atomization) is in
accordance with the work of Roisman et al.~\cite{Roisman2018}. This
transition is mainly related to the evolving thermal and viscous
boundary layers in the spreading drop.

\subsection{Contact and Lift-off Times on Silicon and Sapphire}

The lift-off time $t_l$ of an impacting drop refers to the moment
when the entire drop (including all its visible fragments) retracts
away from the substrate's surface at some velocity, leaving a
clearly visible gap in side views. This corresponds to the usual
'contact' time determination in most of the literature, and its
appearance is similar in optical and X-ray imaging (Here, in X-ray,
the drop size exceeds the field of view.). In a large parameter
range at high temperatures {\color{black}(Regimes III and IV)}, $t_l$
by far
exceeds the actual contact duration $t_c$.\\

\subsubsection{Method and Validation:}

    \begin{figure}[htbp]
        \centering
        \includegraphics[width=\columnwidth]{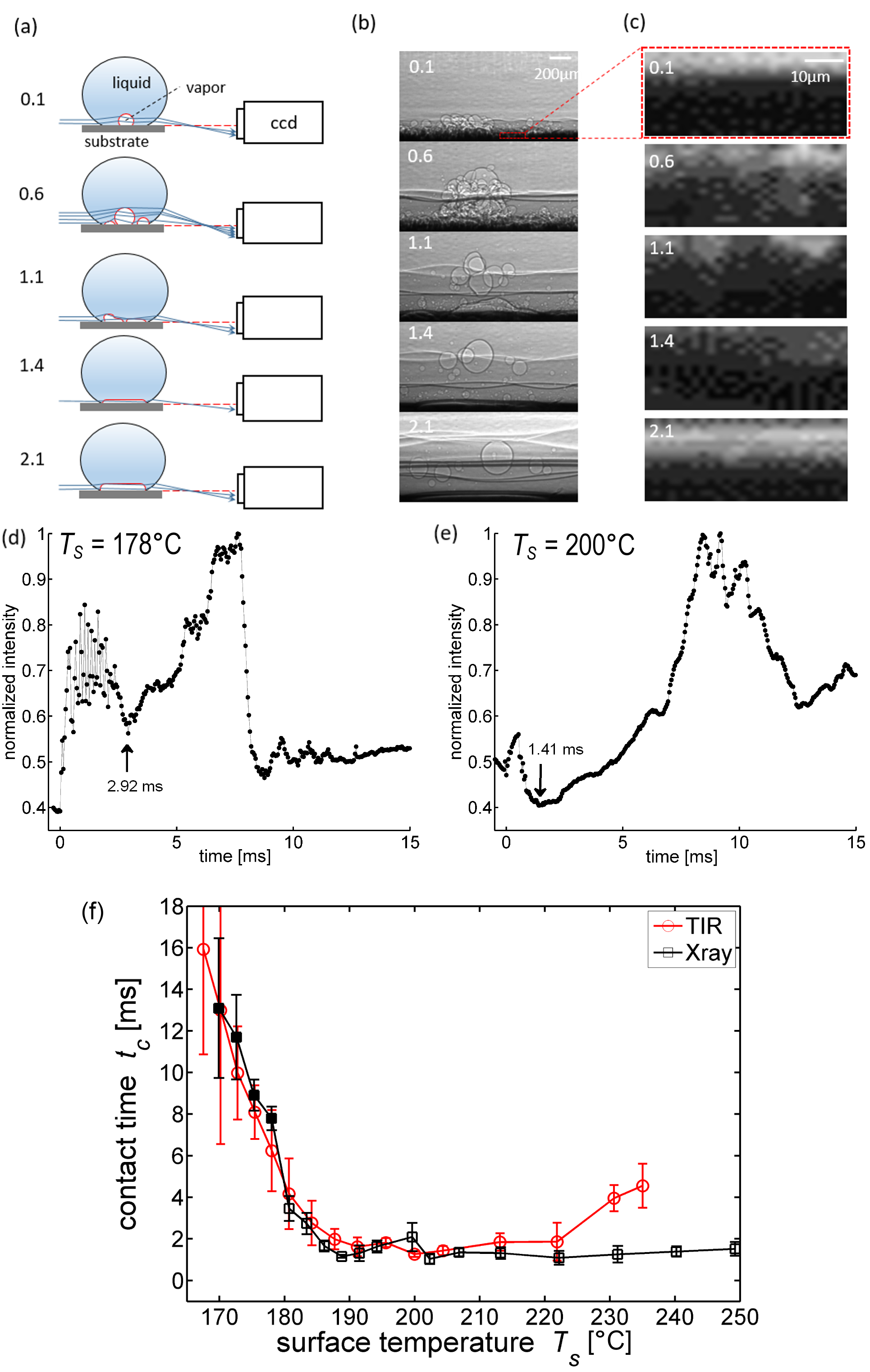}
        \caption{\color{black} Contact time determination from X-ray refraction: (a) schematic, X-ray enters parallel to the substrate's surface and gets partially refracted at corrugations of the phase boundary at the drop underside (blue arrow line). The cumulative refracted beam intensity increases with larger spreading radius of the drop and with larger and more numerous vapour structures, and it decreases as bubbles coalesce and the vapor film flattens. (b,c) Representative X-ray snapshots for (b) the growth and merging of vapor structures and (c) the refracted X-ray beam below the substrate in a small box from (b). (d,e) Exemplary normalized mean intensity evolution from a box of 1 mm width around the impact point, depth $6-46~\rm\mu m$ below the surface, (d) at the onset of Regime III for $T_S=178^{\circ}$C and (e) $T_S=200^{\circ}$C. Arrows mark the local minima corresponding to $t_c$. (f) Comparison of the contact times on sapphire obtained from TIR and X-ray refraction, averages over at least 10 individual experiments, error bars correspond to the standard deviation. X-ray data for $T_S<177^{\circ}$C were obtained from side view (filled symbols), for $T_S>177^{\circ}$C from the refraction (empty symbols). All data:  ethanol on sapphire, $D_D=2~\rm mm$, $U_0=1 ~\rm m/s$.}
\label{fig:X_CTExplanation}
    \end{figure}

Determination of the contact time is obvious in TIR images far from
$T_{L,d}$, see Fig.~\ref{fig:Scenarios}: It corresponds to the
disappearance of all dark (wetted) spots. As TIR can only be applied
to transparent substrates, we develop a new method to extract the
contact time from our phase contrast X-ray data in order to access
contact times on arbitrary smooth surfaces. So far, the literature
investigating impacting drops using X-rays only used the easily
visible absorption contrast between different fluids {\color{black} in transmission} and the slight
refraction of the X-rays at the gas--liquid
interfaces~\cite{Lee2018,Lee2018a,Lee2020,Zhang2011,Zhang2012,Lee2015},
whose interpretation was outlined above and in
Fig.~\ref{fig:setup}(e). {\color{black} Apart from these obvious
features, we observe an X-ray refraction below the substrate's
surface position in the images when the drop's surface is
sufficiently close to the substrate.
Fig.~\ref{fig:X_CTExplanation}(a) shows a sketch of the geometry.
The overall appearance of the refraction-and-absorption induced
image near the impact location is shown in
Fig.~\ref{fig:X_CTExplanation}(b). Images in (c) show the small
contrast-enhanced detail in a small box from (b), right below the
substrate's top surface. We will exploit exactly this refraction to determine the contact
time when it is not directly accessible in side view (mainly Regime III).}

{\color{black} The mechanism is as follows: The X-rays penetrate through the drop and are partly
reflected and partly refracted at the phase boundary between the
liquid and the vapor underneath, which weakly visualizes the
internal vapor structures. The refracted beam can reach below the
substrate as shown by the blue arrows in Fig.~\ref{fig:X_CTExplanation}(a). This increases the
intensity below the substrate's surface locally, see (c,d,e).

For the extraction of the contact time, we only consider mean intensities in a predefined region: First, we determine the impact point and the substrate's top edge as reference point. Next, we sum the intensity for each frame in a fixed region of interest. Our standard choice was a box height of $40~\rm \mu m$
(20 pixel) starting downward at $\approx 6~\rm \mu m$ below the
surface. The width of the box was 1 mm (500 pixel) centered around
the impact point. The box size was accordingly smaller if this
maximal distance exceeded the image size, which occurred due to
off-centered impacts or due to limited frames dimensions at frame
rates exceeding 20889 fps. Next, we determined the mean intensity in
the box for each frame\footnote{Note that the small cut-out in Fig.~\ref{fig:X_CTExplanation}(c) is only illustrative, it is much smaller than the actual box used for the contact time evaluation from mean intensities, e.g., shown in in (d,e).}. Fig.~\ref{fig:X_CTExplanation} (d) and (e)
show two exemplary intensity curves at different temperatures,
normalized to their maxima.}


{\color{black} In correspondence with our TIR results and the side view X-ray images, we interpret the data as follows\footnote{For a slightly tilted substrate, the reflection seen
in Fig.~\ref{fig:setup}(c) displays the same features.}:
The intensity remains spatially and temporally uniform at some reference value while the drop approaches the substrate, see data for $t<0$. As soon as numerous small vapor bubbles and / or channels are formed
due to liquid--substrate contact, the refraction displays spatially inhomogeneous, fluctuating behaviour above those regions where the contacts occur. The mean intensity increases and reaches a first broader maximum (e.g. around 2 ms in (d)). Those local changes in intensity are inevitably connected to the evolution of the boiling patterns underneath the drop, i.e. the intensity changes become smaller and slower with increasing average size of the bubbles, and the
reflection image becomes rather smooth again over regions where the vapor layer has formed and the interface dynamics calm down. The oscillatory behaviour of the intensity before $t = 3~\rm ms$ in (d) is not an artifact, it corresponds to the oscillatory wetting dynamics~\cite{Khavari2017,Harth2019}.
} 




{\color{black} } After that, the intensity decreases to a clear local minimum. This is related to a reduction of the
number and increase of the lateral length scale of the vapour
structures, i.e. merging bubbles and growing vapour channels,
{\color{black} which we likewise observed in TIR. In
Fig.~\ref{fig:X_CTExplanation}(b), one also observes that the mean
height of the small scale vapour structures decreases,
compare the frames at $t=0.6~\rm ms$ and $t=1.1~\rm ms$. Even the large central bubble calms and decreases in height.} 
As a last aspect, it is characteristic for the transition boiling
regime that the contact radius reduces (seen in our TIR data, e.g.
Fig.~\ref{fig:minimum_TIR}, and reported in
Refs.~\cite{Shirota2016,Limbeek2016}), finally producing a
continuous vapor film underneath the entire drop. In this moment,
the vapor film's thickness is assumed to be minimal, resulting in the local minimum of the refraction
intensity marked by the arrows in Fig.~\ref{fig:X_CTExplanation}(d) and (e) after 2.9 ms and 1.4 ms, respectively. 
{\color{black}  The intensity does usually not grow monotonously afterwards, as the vapour layer thickness is inhomogeneous due to the dynamics of the droplet (in contrast to a true Leidenfrost drop in Regime IV). Eventually, the
intensity will decrease again until complete retraction of the drop
from the heated substrate.}

Consequently, we identify the contact time $t_c$ with the occurrence
of the minimum in the local averaged refracted intensity in a large region around the
location of initial impact. {\color{black} Contact times measured from X-ray and TIR, averaged over at
least 10 individual experiments at each temperature, are in excellent agreement in Fig.~\ref{fig:X_CTExplanation}(f). At high temperatures, there is a small delay between impact and the first contact, which is within the error bar. Moreover, contacts are extremely fast, sparse and can be re-entrant, i.e. dissappear completely during a transient period. In the intensity curves, the initial peak can become shallow, however, the local minimum is still visible. This causes the discrepancies between the data sets, and the increase of $t_{c,\rm TIR}$ above $T_S=220^{\circ}$C. X-ray data for $T_S=170^{\circ}$C are unavailable due to the limit of exposure of the scintillator.}

{\color{black} Note that the exact selection of the region under the
substrate's surface is not important for the contact time
determination, in general: Most changes in the box parameters only
affect the relative height of intensity minima and maxima, but not
their moment of occurrence $t$. Changing the height and (within
small range) depth of the box beneath the substrate only changes the
intensity amplitude. The evaluated region must be wide enough to
capture a representative contact region underneath the drop. As
shown in Fig.~\ref{fig:Scenarios}(b,c), also
Fig.~\ref{fig:minimum_TIR}(c), the longest persisting contacts
appear near the impact point, here within approx. 1.5 mm distance
from the location of impact for the analyzed parameters. In
principle, one can average over the full frame width. For large
image widths, this has the disadvantage of averaging the signal too
much. Alternatively, one can analyze the intensity profiles from
several smaller boxes at different locations under the drop, and
take the latest occurrence of the minimum. Too small box sizes cause
disadvantageous signal to noise ratios.

}


\subsubsection{Contact and Lift-off Times:} 
 \begin{figure*}[htbp]
        \centering
        \includegraphics[width=0.75\textwidth]{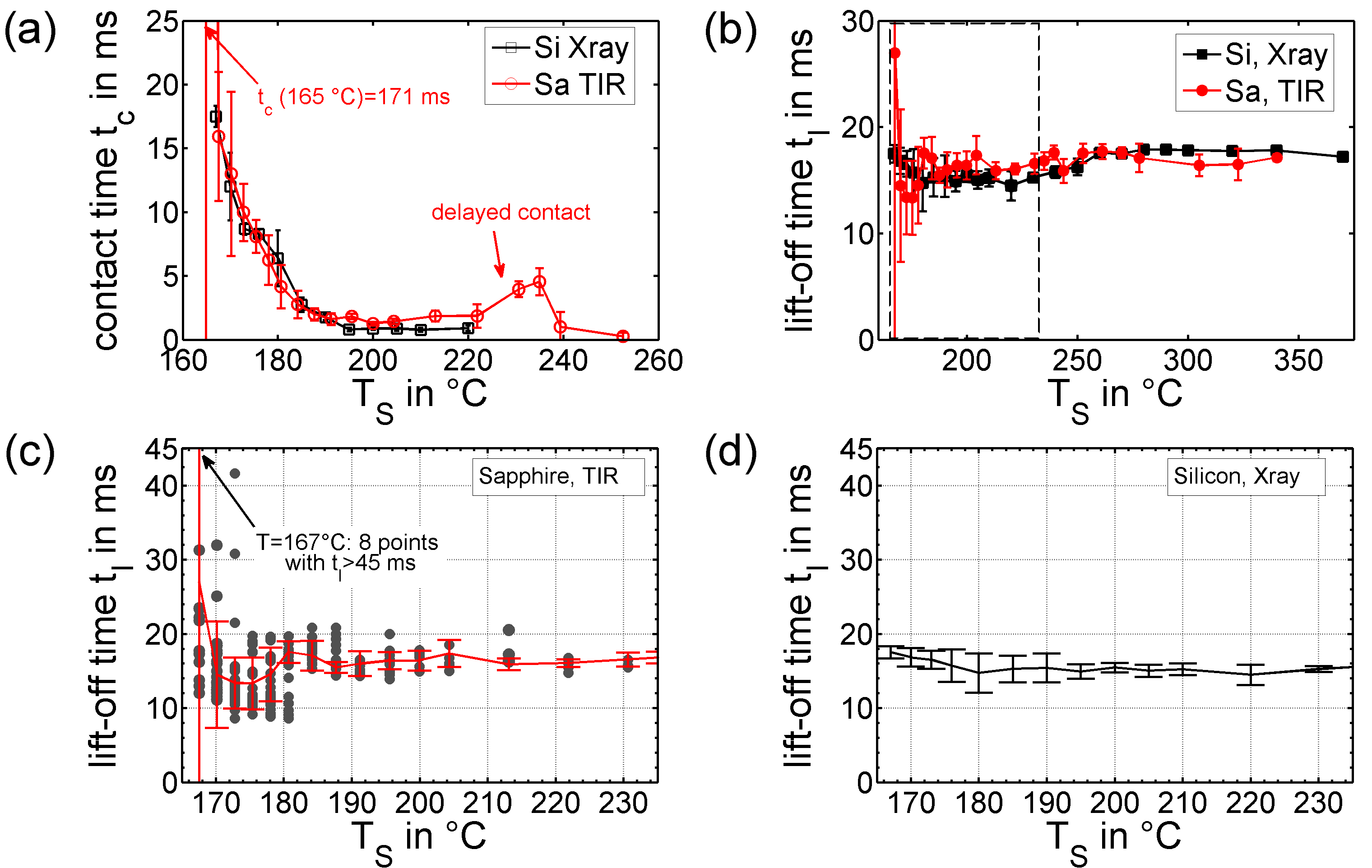}
        \caption{Contact and lift-off times on bulk smooth sapphire (red circles, from TIR) and
        silicon (black squares, from X-ray) in dependence of the surface temperature at impact $T_s$: (a) contact times, (b) lift-off times, (c) and (d) magnify the region in the dashed box in (b) for both substrates. The grey dots in (c) correspond to individual measurements, note that due to the large range of contact times below $T_s=175^{\circ}$C, some data exceed the displayed range and are not shown. Error bars represent the standard deviation of each set of individual impacts under the same conditions. See Fig.~\ref{fig:minimum_statistics} for more details. Points on the curve are average values, error bars indicate the standard deviation of that data set. Ethanol drops of $D_D=2 ~\rm mm$ diameter impacting at velocity $1~\rm m/s$, ${\rm We}=71.2$.\label{fig:ContactTimes_SapphSi}}
    \end{figure*}

        \begin{figure}[ht!]
        \centering
        \includegraphics[width=0.8\columnwidth]{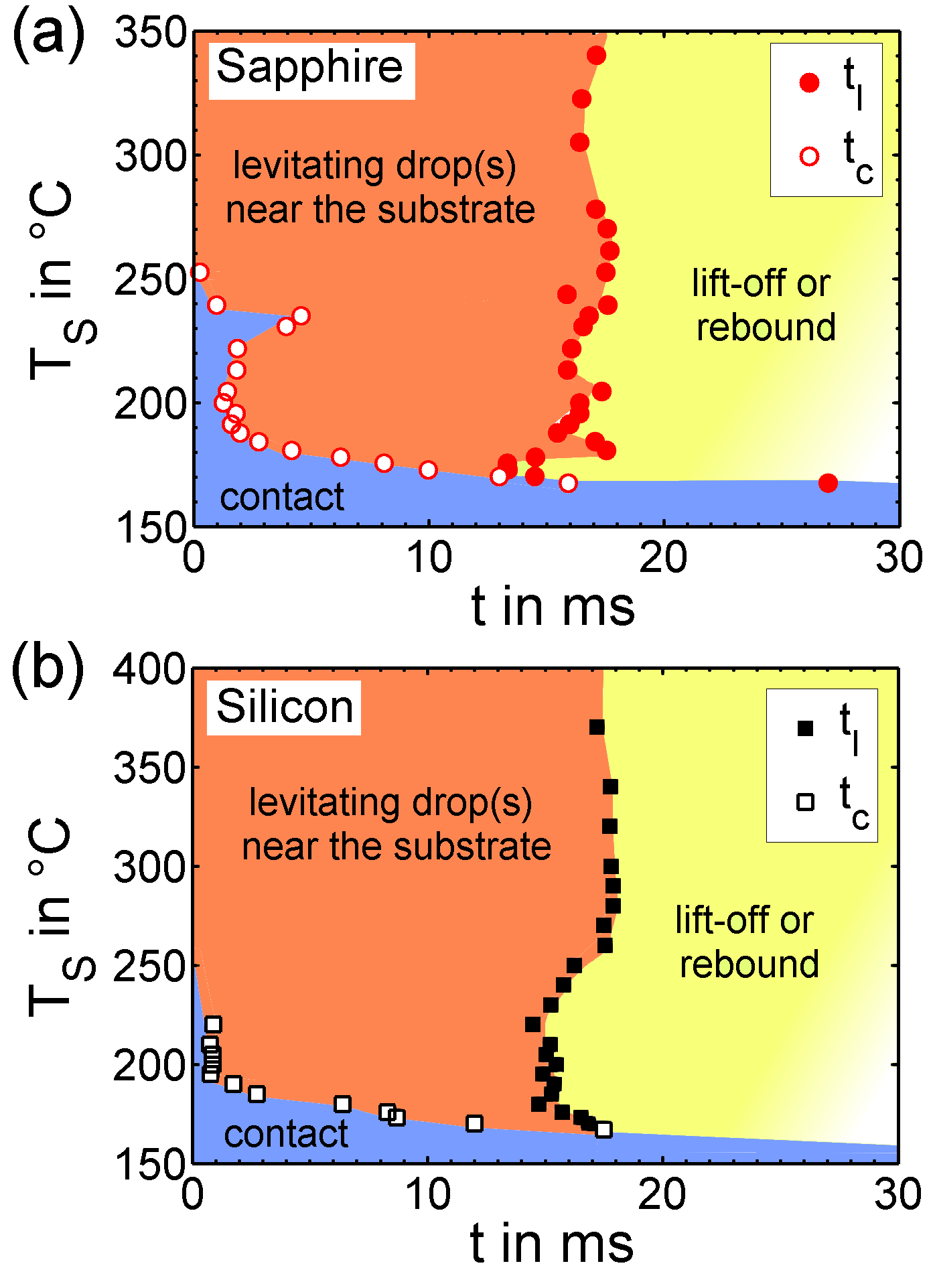}
        \caption{Schematic evolution of boiling characteristics from impact until possible lift-off on
        (a) sapphire and (b) silicon based on the measured mean contact and lift-off times. Interpretation see text.
        Ethanol drops of $D_D=2 ~\rm mm$ diameter impacting at velocity $1~\rm m/s$, ${\rm We}=71.2$.\label{fig:ContactLift-off_ColorGraphs}}
    \end{figure}
    
The lift-off and contact times of ethanol drops
impacting on sapphire and silicon substrates in dependence on the
initial surface temperature are presented in
Fig.~\ref{fig:ContactTimes_SapphSi} over a large range of
temperatures, spanning the four contact and impact regimes described
in the previous section. The observations on both substrates are
very similar. The contact time $t_c$ shown in (a) displays a strong,
continuous decrease between $T_s=160^{\circ}$C and $200^{\circ}$C,
and gradually decreases with further temperature increase. Note that
this is accompanied by drastic changes of the morphology and area
fraction of the global contact as well as duration of the local
contacts, as described in Ref.~\cite{Harth2019}. Above
$T_s=200^{\circ}$C, contacts are very short-lived (lasting few
microseconds) and localized to often sub-micrometer contact sites.
In addition, the initial impact is contact-less, but recurring
contacts appear with some initial delay, and possibly even
intermediate contact-less periods appear at these high temperatures.
This causes the increase in the measured contact times on sapphire
(red curve in (a)) for $T_s > 210^{\circ}$C, as we measured the
total duration between first and last visible contact. Thus, the
determination of the contact time alone will be insufficient to
characterize the heat transfer at high temperatures. Breitenbach et
al.~\cite{Breitenbach2017} reported temperature-independent contact
times of water droplets with a smooth surface in the temperature
range of $200\dots 290^{\circ}$C, where 'thermal atomization' of the
drops occurs, based on top view imaging. This corresponds to the
high-temperature range of Regime I, and II in our measurements. They
developed an empirical model, where basically the duration of
contact corresponds to the moment when the thickness of the thermal
boundary layer reaches the thickness of the flattened
lamella~\cite{Roisman2018}, so that
\begin{eqnarray}
 t_c=\frac{0.6 D_D^2}{b^2\alpha_l {\rm Re_0}^{0.8}}\left(\frac{\nu}{\nu_0}\right)^{0.8}, \label{eq:tc_Roisman2018}
\end{eqnarray}
where the $\alpha_{l}=k_l/(\rho c_{p})$ is the thermal diffusivity,
$k$ is the thermal conductivity,
$\rho$ the density and $c_p$ the isobaric specific heat capacity, and $\nu$ is the kinematic viscosity of the
liquid at the specific superheat temperature.
The Reynolds number $\rm Re_0$ and the kinematic viscosity $\nu_0$ are
taken at room temperature. From data taken at different Weber
numbers, Roisman et al.~\cite{Roisman2018} could deduce that the constant $b=1.0$
for water. The model's prediction agrees very well with the data
given in Ref.~\cite{Roisman2018} for water.
 Comparing to our contact times of ethanol drops, Fig.~\ref{fig:ContactTimes_SapphSi}(a),
 no range of (even roughly) constant $t_c$ for $T_s<200^{\circ}C$ exists, in {\em qualitative} disagreement to Ref.~\cite{Roisman2018}.
 One might at first assume that this may be due to changing material parameters with increasing temperature, but this
 is not the case: Inserting the temperature-dependent thermal diffusivity and kinematic viscosity of ethanol into Eqn.~(\ref{eq:tc_Roisman2018}),
 based on the thermal conductivity $k$ calculation from Assael et al.~\cite{Assael2013} and the other temperature-dependent material parameters given in the DDBST Dortmund data base (e.g. Refs.~\cite{EthDynViscosity,EthDensity}),
  and using $b=1.0$, we obtain contact times $t_c$ between $7.5$ and $7 ~\rm ms$ with a decreasing trend in the range between $T_s=165^{\circ}$C and $200^{\circ}$C.
  This order of magnitude is agreeable below $180^{\circ}$C, and changing $b$ does not change the qualitative trend.
  Again, the method of contact time determination may play an additional role: Our measured optimal lift-off times of
  $t_{l,\rm min}\approx~\rm 8~\rm ms$, Fig.~\ref{fig:ContactTimes_SapphSi}(c), agree well with Roisman's estimation providing the time of lamella rupture.
  After rupture, the lamella retraction is expected to occur analogue to the rupture of low-viscosity liquid films presented by Taylor~\cite{Taylor1959} and
  Culick~\cite{Culick1960}, see also Ref.~\cite{Villermaux2007}.\\

 Let us now analyze the duration that the drops reside in vicinity of the substrate, the lift-off times $t_l$,
 Fig.~\ref{fig:ContactTimes_SapphSi}(b)--(d). At first glance, again the data on silicon and sapphire are similar and agree
 within the standard deviation of the data sets at given substrate temperature $T_s$. Contact and lift-off times in Regime I
 are on the order of few a seconds (not shown in the plot). Thus, the better thermal conductivity of silicon compared to sapphire has only small influence on the contact and lift-off times here. For impact on silicon (black squares, in (b) and (d)), we observe an initial slight decrease of the lift-off time below $T_s=180^{\circ}$C, and an almost constant lift-off time for $180^{\circ}$C $\leq T_s\leq 240^{\circ}$C. Above $T_s=240^{\circ}$C, $t_l$ slightly increases, followed by a slight decrease toward higher temperatures
 in the Leidenfrost Regime IV, continuing until $T_s=550^{\circ}$C, where $t_l\approx 11.6~\rm ms$. This slow decrease at high temperatures can be expected given that the vapor layer separating drop
 and substrate grows in thickness~\cite{Lee2018a}. Thus, the small friction in the vapor layer decreases with increasing temperature,
 an one may presupposes that less surface energy is dissipated.
 Between $T_s=165^{\circ}$C and $240^{\circ}$C, $t_{l,\rm Si}$ is roughly constant between $16~\rm ms$ and $18~\rm ms$.
 This experimentally determined rebound time agrees with the expectation from literature: There, it is approximated by the period of
 an oscillating drop~\cite{Makino1984} as $t_l \approx \pi/4\sqrt{\rho_0 D_D^3/\gamma} = 13.3~\rm
 ms$, using the material properties at room temperature. Slightly
 larger empirical pre-factors than $\pi/4$ were given by Biance et
 al.~\cite{Biance2006} (prefactor 0.937, $t_l\approx 15.8~\rm ms$)
 and Chen et al.~\cite{Chen2007} (prefactor 1.12, $t_l\approx 18.9~\rm
 ms$).

 The lift-off times on the sapphire substrates display a shallow local minimum between $T_s=170^{\circ}$C and $180^{\circ}$C, where the
 mean lift-off time reduces to only $t_{l,~\rm Sa}\approx 13.3~\rm ms$, before it rises again for higher temperatures to $16~\rm ms$ and $18~\rm ms$
 (the same value as measured on Si). In the enlarged plot of
Fig.~\ref{fig:ContactTimes_SapphSi}(c), we added the values in the individual measurements. One observes that numerous
points center around $\approx 10~\rm ms$, whereas measurements with much
higher lift-off times, occur at the same substrate temperature. More
details and the mechanism behind this substantial lift-off time
reduction are provided in
 Sec.~\ref{Sec:Holes}.\\

 Last, let us sketch the general time evolution of the contact characteristics of a drop impacting on a hot plate in terms of Fig.~\ref{fig:ContactLift-off_ColorGraphs}: Time evolves along the ordinate axis. The colored regions in the plots correspond to certain contact behavior: the blue region denotes contact with the substrate. In the red regions, the droplet is close to the substrate but contact-less (e.g. a Leidenfrost drop during spreading and receding), and the yellow region corresponds to times when the drop has accomplished its first lift-off or rebound. At low temperatures, the lift-off velocity is small and the re-approach of the droplet to the substrate occurs after very short time.

 An ethanol drop impacting on a substrate at $T_s=150^{\circ}$C will remain in contact (blue region)
 all the time until it fully decomposed or evaporated (Regime I). At slightly higher temperature, $t_c=t_l$ (Regime II),
 i.e. the drop is initially attached to the substrate, but finally lifts off the substrate (direct transition from the blue to the yellow region).
 The width of this temperature range is only around 10~K (cf. also Fig.~\ref{fig:StateDiagram}). For a broad range of
 temperatures, the contact is lost prior to lift-off or rebound (Regime III, range $170^{\circ}$C$\leq T_s\leq 250^{\circ}$C on sapphire).
 This refers to a first cross-over from blue to orange, and then to the yellow region. Finally, in the Leidenfrost regime IV, drops spread and recede close to the substrate without any contact, before they rebound (transition red to yellow).

\section{Droplet Dynamics near the Minimum Lift-off Time}


\label{Sec:Holes}
    \begin{figure}[h]
        \centering
        \includegraphics[width=\columnwidth]{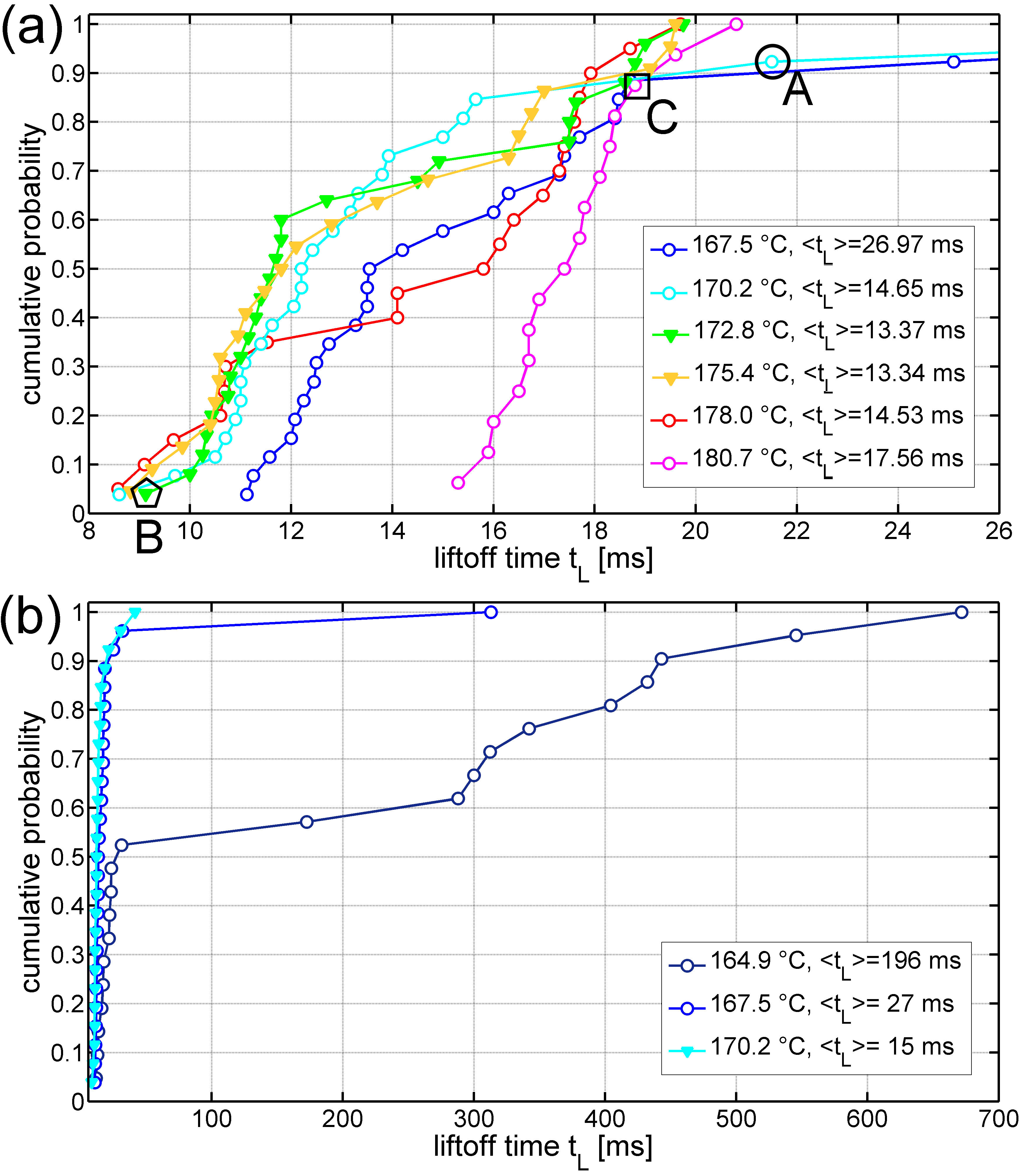}
        \caption{Statistics of lift-off times in Regime II: Cumulative probability
        distributions of the lift-off time and mean lift-off times in dependence of the initial substrate temperature.
        The minimum occurs around $173\dots 175^{\circ}$C (green and orange curves, triangles), with average $t_{l,~\rm min}\approx13.3~\rm ms$. The probability distributions below these temperatures possess 2 peaks, one at low, and one at high temperatures, seen by the drastic change of slope in the blue curves (b). Marked points in (a) correspond to image sequences in Fig.~\ref{fig:minimum_imageSeq}. Ethanol drops of $D_D=2~\rm mm$ diameter impacting at velocity $1~\rm m/s$ on the sapphire substrate, measurements from side view imaging.} \label{fig:minimum_statistics}
    \end{figure}

    \begin{figure}[ht!]
        \centering
        \includegraphics[width=\columnwidth]{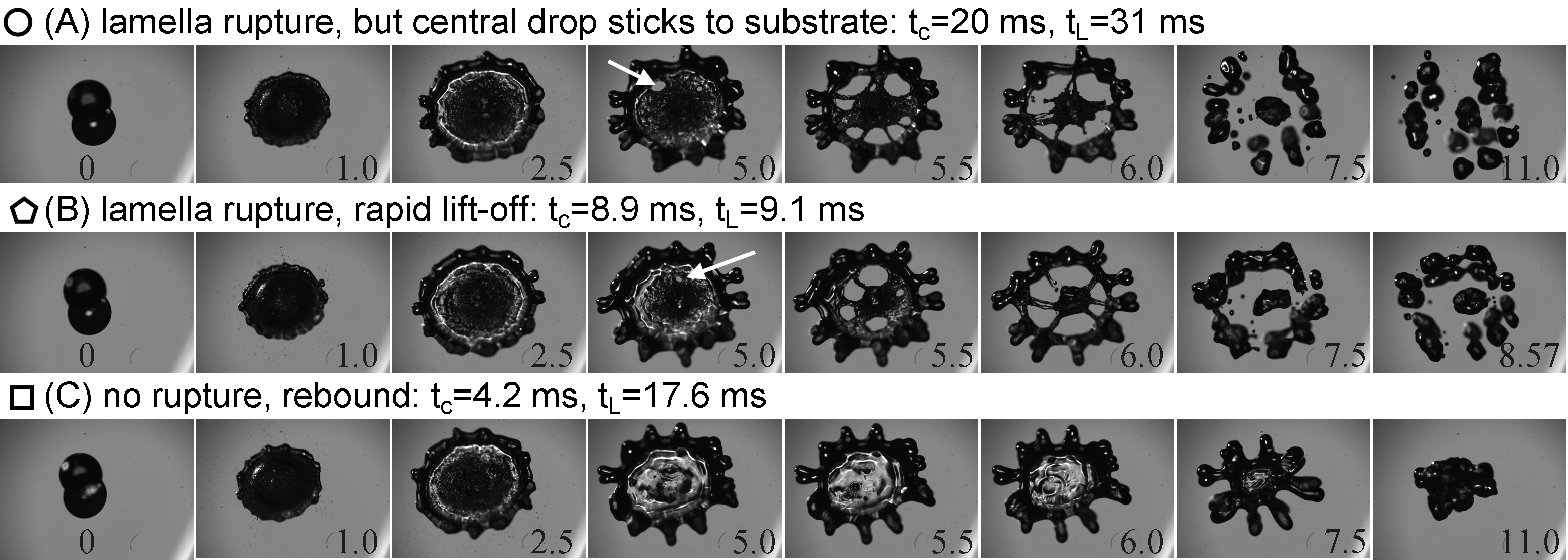}
        \caption{Hole formation and sticking govern the lift-off, exemplary tilted top view image sequences corresponding
        to the marked points in Fig.~\ref{fig:minimum_statistics}(A): White arrows indicate early formed holes. (a) Lamella rupture with insufficiently fast retraction of the contact underneath the central drop region, central droplet is stuck, practically no lift-off. (B) optimal situations: hole expansion and sufficiently fast retraction of the contact (see also Fig.~\ref{fig:minimum_TIR}), lift-off in shape of a disintegrated pancake. (C) Increased temperature with rebound of a single drop, Regime III, heat transfer during spreading is insufficient to cause lamella rupture, $t_c \ll t_l$. The arrows in (A) and (B) indicate the first hole formed in the lamella. Numbers give the time after impact (identified in TIR) in milliseconds. Maximum spreading is reached after $\approx 6.35~\rm ms$. Ethanol drops of $D_D=2~\rm mm$ diameter impacting at velocity $1~\rm m/s$ on the sapphire substrate. \label{fig:minimum_imageSeq}}
    \end{figure}

        \begin{figure}[ht!]
        \centering
        \includegraphics[width=\columnwidth]{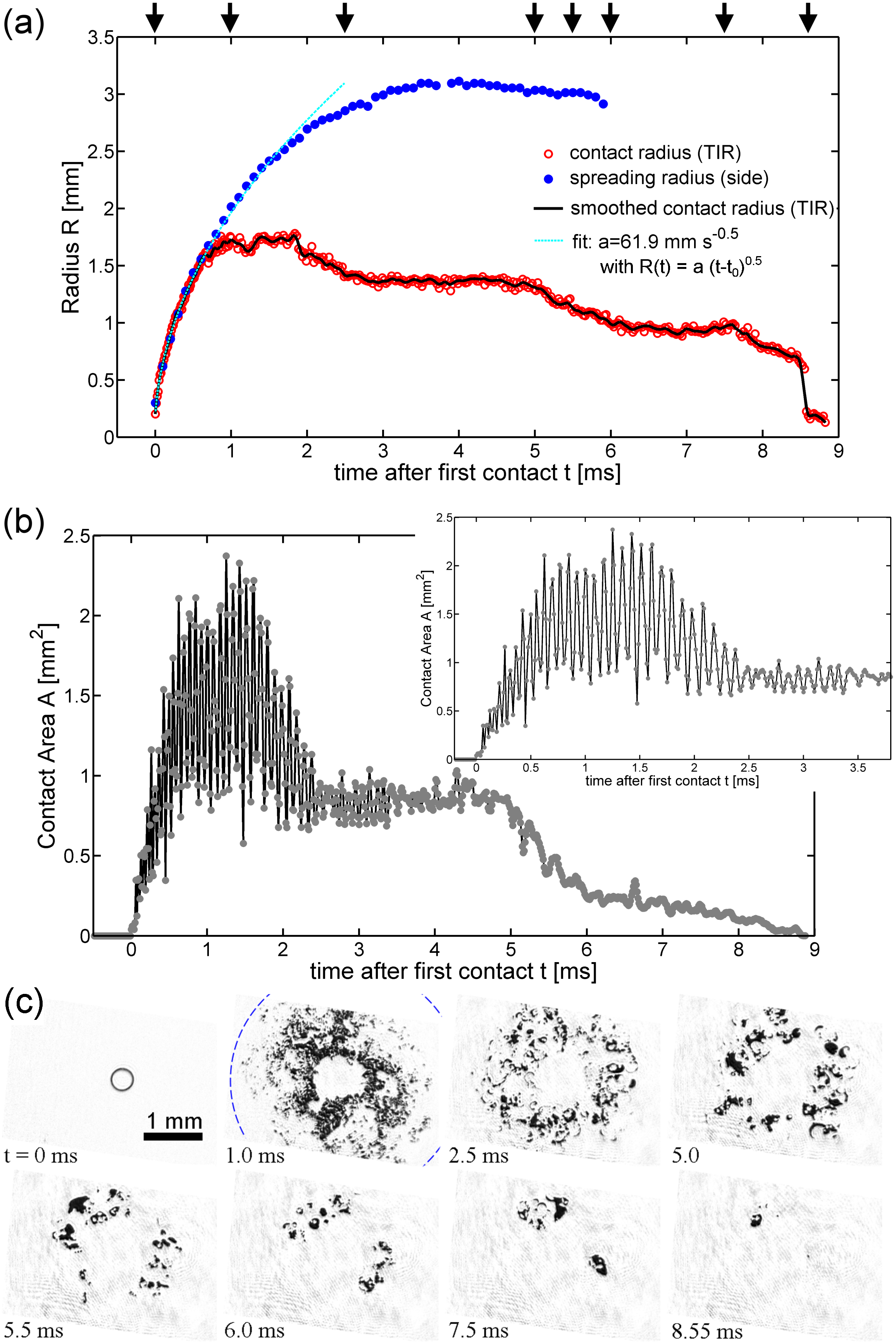}
        \caption{Contact, hole formation and spreading dynamics for a drop with small lift-off time in the minimum
        region in Regime II, see Fig.~\ref{fig:minimum_imageSeq}(b): Ethanol drop of $D_D=2~\rm mm$, $U_0=1~\rm m/s$,
        impacting on a sapphire substrate at initial temperature $T_{\rm s}=175.4^{\circ}$C. (a) spreading radius, fit with $R(t)=a\sqrt{t-t_0}$,
        with $t_0$ the moment corresponding to zero contact radius, and $a=61.9~\rm mm\, s^{-0.5}$. (b) Time dependence of the contact area measured by
        thresholding the TIR images: the contact and lift-off times approximately coincide, $t_c=8.9~\rm ms$ and $t_l=9.1~\rm ms$. (c) TIR images showing the
        contact morphology, corresponding to the times marked by arrows in (a), and to the tilted top view image sequence
        Fig.~\ref{fig:minimum_imageSeq}(B). The dashed circle at $t=1~\rm ms$ indicates the contact radius in side view.
        At all later times, it by far exeeds the field of view of the TIR data. }
        \label{fig:minimum_TIR}
    \end{figure}

In order to elucidate the origin of the lift-off time reduction on sapphire for surface temperatures between $T_s=170\dots 180^{\circ}$C, we performed $\approx$ 25 separate measurements for each temperature, and augmented the side and bottom TIR images with videos taken in a tilted top view perspective (see setup sketch in Fig.~\ref{fig:setup}(b)). In this lift-off time reduction range, which makes up most of Regime II, droplets fragment into smaller drops, and lift off the substrate at comparatively low velocity in a flattened, disintegrated overall shape, see Figs.~\ref{fig:Scenarios}(b),~\ref{fig:ScenariosSketch}(b). The transition from contact boiling (Regime I) to lift-off (Regime II) is not sharp. The two typical impact scenarios are sketched in Fig.~\ref{fig:ScenariosSketch}(b):

In the lower temperature range of Regime II, we find individual
impact events with long contact and lift-off times (tens to hundreds of milliseconds, not all data shown in Fig.~\ref{fig:ContactTimes_SapphSi}) as well as such with very short ones
(less than 10 ms). In Figure~\ref{fig:minimum_statistics}, we show
the cumulative probability distributions of the lift-off times $t_l$
for different initial substrate temperatures $T_s$: These
distributions are usually non-Gaussian, often showing a population
of impact events with short lift-off times and one with larger
values. The mean \emph{contact} time continuously decreases in this
range, cf. Fig.~\ref{fig:ContactTimes_SapphSi}(a). For $T_s\leq
170.2^{\circ}$C, we find an increasing number of impact events with
$t_l<20~\rm ms$, coexisting with impacts with very long contact and
lift-off times of up to $0.7~\rm s$, see
Fig.~\ref{fig:minimum_statistics}(b).
At high temperatures, $T_s>180^{\circ}$C, in Regime III, lift-off
times display a narrow distribution between $15\dots 18~\rm ms$.
Drops rebound without disintegration, and this range coincides with
expectations based on inertial, contact-less rebound. In the range
of $170^{\circ}$C$<T_s<180^{\circ}$C, numerous impacts display a
reduced lift-off time by up to a factor of $\approx 2$, to typically
$t_l=7.8~\rm ms$ and $11~\rm ms$. The minimal mean lift-off time
occurs around $T_s=(174\pm 1)^{\circ}$C, see green and orange curves
in Fig.~\ref{fig:minimum_statistics}(a).

 Tilted top view imaging
indicates the mechanism behind this drastic contact time reduction,
see Fig.~\ref{fig:minimum_imageSeq}: Lamella rupture causes droplet
disintegration in a spread-out state, see (A) and (B). Such thermal
atomization of a drop on a hot plate occurs when the temperature at
the top of the spread-out lamella reaches the boiling point of the
liquid~\cite{Roisman2018}, and bubbles rupture through the liquid
film. However, this alone is insufficient to cause the lift-off time
reduction: Additionally, the lamella retraction must be easily
possible, i.e. the contact between the substrate and the drop must
recede sufficiently fast to avoid contact line pinning, as in (B).
{\color{black} Only when both conditions are fulfilled}, the lift-off
time is substantially reduced.\\

 Drop spreading / receding and contact dynamics in the situation of  a short lift-off time are exemplarily
 analyzed in Fig.~\ref{fig:minimum_TIR}, corresponding to Fig.~\ref{fig:minimum_imageSeq}(B) and the Regime II
 in Fig.~\ref{fig:ScenariosSketch}(b): The temporal evolution of the spreading radius (blue) and the radius of
 the smallest circle containing all contact spots (red) is shown in (a). For $t\geq 0.66~\rm ms$, the contact radius
 is smaller than the spreading radius. The outer regions of the spreading lamella levitate, while the central regions
 remain in contact. This is characteristic for transition boiling of impacting drops~\cite{Shirota2016,Limbeek2016}. Drop spreading
 follows the $R(t)\propto\sqrt{t-t_0}$ scaling rigorously derived by Riboux and Gordillo~\cite{Riboux2014}.

 For $t>0.66~\rm ms$, the contact radius first remains roughly constant and then slowly recedes. The contact area, see Fig.~\ref{fig:minimum_TIR} (b), displays high-frequency oscillations for $t\leq 3~\rm ms$. This phenomenon is described in Refs.~\cite{Khavari2017,Harth2019}. After that, the contact area decreases more rapidly. In particular, contacts group into several smaller spots instead of being equally distributed (cf. image sequence in (c)). This particular drop lifts off the substrate at $t_l\approx9.1~\rm ms$. Impact events with longer lift-off times at similar temperatures correspond to longer lasting contact of the central fragment of the drop, remaining stuck to the substrate after lamella rupture caused fragmentation, as e.g. for Fig.~\ref{fig:minimum_imageSeq}(A). It is always this central fragment of the initial drop which avoids the lift-off. Fragments resulting from breakup of the rim of the lamella
 levitate.\\


 In the present experiments, the first holes in the lamella appear due to a growing vapor bubble while the drop is still spreading, indicated by arrows in Figs.~\ref{fig:minimum_imageSeq}(A) and (B). This process requires an efficient heat transfer from the substrate to the drop liquid, which only occurs while direct contact between the drop and the liquid persists, i.e. the levitating parts of the lamella can be assumed to heat up more slowly than the parts which are in contact with the substrate. This sets an upper limit to the regime of thermal atomization, which is caused by vapor bubbles rupturing through the lamella. Thus, when a large part of the lamella levitates already early during the spreading process, the lamella remains intact and the drop will rebound from the hot surface as a whole (see Fig.~\ref{fig:minimum_imageSeq}(C)).\\

 Finally, we note that we have not observed a similar phenomenon for ethanol drop impact on silicon, where only X-ray data is available. Close inspection of the X-ray data did not indicate any lamella rupture, although otherwise the drop interaction with that substrate was very similar to the sapphire disks. More detailed studies would be needed to reveal the reason for this discrepancy, which could be related to heat transfer properties but also to changed wettability and thus contact line behavior of the vapor bubbles.

    \section{Discussion and Conclusions}
\label{Sec:Conclusions}

    We combine Ultrafast synchrotron phase contrast X-ray, high-speed Total Internal Reflection
    and conventional side and top-view imaging to study the contact and rebound of ethanol drops
    impacting on smooth hot plates. Exploiting the weak reflection and refraction of X-rays on topographic structures
    on the underside of a boiling drop, we developed a new method of reliable contact detection
    applicable to droplets impacting on hot non-transparent substrates such as metallic plates.
    The method was validated by direct comparison of contact time measurements on sapphire between
    X-ray and (optical) Total Internal Reflection data, with excellent agreement. This finally
    allows us to avoid the ambiguities of typical contact time measurements from side and top
    view data, and it allows a direct quantitative comparison of the mean contact times on sapphire
    and the even better heat conductor silicon.

    The contact times display a monotonous decrease
    with increasing temperature, and agree within error bars for both substrates except near the
    dynamic Leidenfrost point. At such high temperatures, contact usually
    occurs with a delay respective to the moment of impact. This may be interpreted as an effect of
    rapid substrate cooling due to the interaction with the colder, evaporating drop similar to van
    Limbeek's results regarding (badly heat conducting) glass substrates~\cite{Limbeek2016}. Using
    our combined method, we identify localized, short-lasting contacts at substrate temperatures more
    than 50~K above the previously determined Leidenfrost transition (based solely on the initial
    impact and TIR intensity)~\cite{Shirota2016,Khavari2017}.

    Drop lift-off from the heated substrate is observed also at substrate temperatures, where
    substantial contact with the substrate exists. To analyze this phenomenon more deeply, we
    measure the lift-off times of the droplets in side view. Impact events are classified into the
    four regimes sketched in Fig.~\ref{fig:ScenariosSketch}, based on the occurrence and duration of
    contact and until lift-off. Between the classical contact and Leidenfrost impact regimes, we
    observe a large range of higher temperatures where droplets rebound from the hot substrate after
    initial contact. Here, the contact times are almost constant, slightly decreasing with increasing
    temperature, between $16~\rm ms$ and $20~\rm ms$ both on sapphire and silicon. The drops rebound in increasingly
    prolate shapes with increasing velocities for increasing substrate temperatures. A small range of
    10 K to 15 K at low temperatures with approximately equal the contact and lift-off times of the
    drop exists. This regime coincides with the appearance of the thermal atomization regime~\cite{Roisman2018}.
    However, the measured contact times qualitatively disagree with the model's predictions~\cite{Roisman2018}:
    The model predicts a constant contact time of $\approx 8~\rm ms$, which is the correct order of magnitude,
    but the constancy of contact time is in sharp disagreement with our measurements. The discrepancy is not
    straightforwardly explained by a change of material parameters. Drops in this regime fragment into on the
    order of 10 smaller droplets, and lift off the substrate at small velocity in the shape of a disintegrated
    pancake. Only the transition between the short-lived contact and the Leidenfrost (film boiling) regimes is
    strongly temperature dependent.

     Strikingly, on the sapphire substrate, the mean lift-off times of those droplets where contact and
     lift-off times are approximately equal is substantially smaller than at higher temperatures, even
     than that of Leidenfrost drops. The statistics of lift-off times in individual impact events in this
     temperature range display a non-Gaussian distribution around the mean, with typically one subset of
     points at small lift-off times between 8 and 11 ms and a large range of other data for up to few hundreds
     of milliseconds. We identify an optimal timing of spontaneous lamella rupture caused by bursting vapor
     bubbles with the receding of the contact underneath the drop as the physical mechanism behind the contact
     time reduction by up to a factor of 2 (down to $\approx 8~\rm ms$). At low temperatures, this regime is
     limited by persisting contacts underneath the spreading and receding lamella, causing strong contact line
     pinning after lamella rupture events. The upper temperature limit is set by the aspect that sufficiently
     large and lasting contacts are required for an efficient heat transfer. This is the precondition for the
     thermal boundary layer in the liquid reaching the top of the lamella, which sets the time of thermal
     atomization~\cite{Roisman2018}. The transition boiling regime of drop impact on hot plates is characterized
     by the fact that an increasing outer part of the lamella levitates on vapor at earlier times with increasing
     plate temperature~\cite{Shirota2016}. This reduces the heat transfer efficiency, such that lamella rupture
     is absent above a certain temperature; thus the rupture-based mechanism of contact time reduction is eliminated.

     Similar contact time reduction and rebound of droplets in a spread-out  (pancake) shape was reported on
     superhydrophobic substrates~\cite{Chantelot2018,Liu2014}. Levitating drops over hot plates (with neglect
     of the vapor flow) are frequently seen as an ideal realization of super-hydrophobicity, as their advancing
      and receding contact angles equal $180^{\circ}$. Chantelot et al.~\cite{Chantelot2018} studied water droplets
      impacting on a super-hydrophobic substrate with a topographic defect. By this method, they cause lamella rupture
      in a controlled position and time, and predict the contact time reduction using a simple model. For our drops
      impacting on a hot plate, lamella rupture is less controlled, and occurs at several positions at different times.
      The advancing contact angle is always effectively $180^{\circ}$ (except for the contact boiling regime), as the
      lamella levitates on a vapor layer. The resistance against receding of the drop upon lamella retraction is
      governed by the boiling-related receding of the drop's contact with the hot plate: The receding contact angle
      is large where the lamella levitates, but it is small (similar to the ethanol--sapphire contact angle) where
      contact persists.\\

 Our study explicitly shows that the transparent sapphire substrates used in many optical experiments provide
 an equally good basis for the measurements of contact times as heated metal substrates. An exception may apply for
 water drops, due to their much larger latent heat of vaporization. We provided reliable data for the contact times,
 which significantly differ from 'contact' times in the literature, which are typically determined from conventional
 side and top view imaging. Contact times monotonously decrease with increasing temperature, and micrometer-sized
 localized contacts may last for only microseconds when approaching the dynamic Leidenfrost transition. Neither the
 absence of spray nor the disappearance of structures on the spreading lamella provide a reliable criterion for contact
 determination. Our data can form an initial step towards more detailed experimental studies and towards a reliable basis
 for developments of theoretical models of the coupled boiling-and impact dynamics of drops on hot plates. Details of the
 mechanism of drop fragmentation and rebound need to be analyzed. This mechanism can be potentially exploited as a route
 towards controlled droplet removal, e.g.  by smart design of structured surfaces.

\section*{Conflicts of interest}
There are no conflicts to declare.

\section*{Acknowledgements}
   The authors would like to acknowledge the staff at the 32-ID beamline at Argonne National Laboratory for their support in realizing the
    X-ray measurements. The use of the Advanced Photon Source, an Office of Science User Facility operated for the U.S. Department of Energy (DOE)
    Office of Science by the Argonne National Laboratory, was supported by the U.S. DOE under contract no. DE-AC02-06CH11357.
    S. H. L, M. K. and J. H. J. acknowledge support by the
    National Research Foundation of Korea (NRF) grant funded by the Korean government
    (NRF-2017R1E1A1A01075274) and Brain Korea 21 PLUS Project for the Center for Creative Industrial Materials.
    K. H. acknowledges the German Science Foundation (Deutsche Forschungsgemeinschaft, DFG) for funding within
    Grants HA-8467/1 and HA-8467/2--1. D. L. acknowledges funding by ERC grant 740479-DDD. K. H., M. R. and D. L. thank the Max Planck Center Twente for funding.



\balance



\providecommand*{\mcitethebibliography}{\thebibliography}
\csname @ifundefined\endcsname{endmcitethebibliography}
{\let\endmcitethebibliography\endthebibliography}{}

\end{document}